\documentclass[preprint,12pt]{article}

\usepackage[numbers]{natbib} 
\usepackage[table]{xcolor}
\usepackage{enumitem}
\usepackage{tabularx}
\usepackage{ragged2e}
\usepackage{booktabs}       
\usepackage{amsfonts}       
\usepackage{nicefrac}       
\usepackage{microtype}      
\usepackage{hyperref}
\usepackage{graphicx}
\usepackage{amssymb}

\usepackage{lineno}
\usepackage{arxiv}
\usepackage{amsmath}
\usepackage{amsfonts}
\usepackage{subcaption}
\usepackage{physics}

\usepackage{tikz}
\usetikzlibrary{arrows.meta}
\usetikzlibrary{tikzmark, positioning, fit, shapes.misc, shadows}
\usetikzlibrary{arrows}
\usetikzlibrary{decorations.pathreplacing, calc,shapes.geometric}

\tikzset{brace/.style={decorate, decoration={brace}},
  brace mirrored/.style={decorate, decoration={brace,mirror}},
}
\usepackage{algpseudocode,algorithm,algorithmicx}

\makeatletter
\def\ALG@special@indent{%
    \ifdim\ALG@thistlm=0pt\relax
        \hskip-\leftmargin
    \else
        \hskip\ALG@thistlm
    \fi
}
\newcommand{\Begin}[1]{\item[]\noindent\ALG@special@indent \textbf{Begin:}\ #1}
\newcommand{\Establish}[1]{\item[]\noindent\ALG@special@indent \textbf{Establish:}\ #1}
\newcommand{\End}{\item[]\noindent\ALG@special@indent \textbf{End}}
\makeatother
\algblock[Name]{Start}{End}
\algblockdefx[NAME]{START}{END}%
    [2][Unknown]{Start #1(#2)}%
    {Ending}

\usepackage{nccmath}

\usepackage{xcolor}
\usepackage[framemethod=TikZ]{mdframed}
\definecolor{ikmgray}{HTML}{5E5E5E}
\definecolor{ikmgreen}{HTML}{C9DA2B}
\newmdenv[frametitle={},
middlelinecolor=ikmgreen,
middlelinewidth=0pt,
backgroundcolor=ikmgray!20,
roundcorner=2pt,
bottomline=false,
leftline=true,
topline=false,
rightline=false,
skipabove=10pt,
skipbelow=10pt,
leftmargin=10pt,
rightmargin=10pt,
innerleftmargin=5pt,
innerrightmargin=10pt,
innertopmargin=10pt,
innerbottommargin=10pt]{Algorithmus}
\usepackage[labelfont=bf, 
format=plain, 
labelsep=endash, 
justification=raggedright 
]{caption}
\DeclareCaptionType{kasten}[Box]
\usepackage{bm}

    \DeclareFontFamily{OT1}{pzc}{}
\DeclareFontShape{OT1}{pzc}{m}{it}{<-> s * [1.10] pzcmi7t}{}
\DeclareMathAlphabet{\mathpzc}{OT1}{pzc}{m}{it}
\title{A review on data-driven constitutive laws for solids}
\author{
Jan Niklas Fuhg \\
Department of Aerospace Engineering and Engineering Mechanics \\
The University of Texas at Austin,
Texas, USA \\
\And
Govinda Anantha Padmanabha  \&
Nikolaos Bouklas  \\
Sibley School of Mechanical and Aerospace Engineering \\
Cornell University, 
New York, USA \\
\And
Bahador Bahmani \&   WaiChing Sun \\
Department of Civil Engineering and Engineering Mechanics\\
Columbia University,
New York, USA
\And
Nikolaos N. Vlassis\\
Department of Mechanical and Aerospace Engineering\\
Rutgers University,
New Jersey, USA
\And
Moritz Flaschel, Pietro Carrara \& Laura De Lorenzis \\
Department of Mechanical and Process Engineering \\
ETH Zürich,
Zürich, Switzerland
}

\begin{document}
\setlength{\parindent}{0cm}
\maketitle


\begin{abstract}
This review article highlights state-of-the-art data-driven techniques to discover, encode, surrogate, or emulate constitutive laws that describe the path-independent and path-dependent response of solids. 
Our objective is to provide an organized taxonomy to a large spectrum of methodologies developed in the past decades and to discuss the benefits and drawbacks of the various techniques for interpreting and forecasting mechanics behavior across different scales. Distinguishing between machine-learning-based and model-free methods, we further categorize approaches based on their interpretability and on their learning process/type of required data, while discussing the key problems of generalization and trustworthiness. We attempt to provide a road map of how these can be reconciled in a data-availability-aware context. We also touch upon relevant aspects such as data sampling techniques, design of experiment, verification, and validation.
\end{abstract}


\section{Introduction}
Problems in solid mechanics are formulated in terms of three sets of equations: the first encodes basic conservation principles (e.g., balance of linear momentum) and governs the equilibrium of deformable bodies following the definition of a stress tensor; the second describes the kinematics of motion in terms of displacements, strains and strain rates. The third set, denoted as \textit{constitutive law} (CL) (or \textit{material model}), describes the response of a material to external stimuli by establishing a relation between kinematic and static quantities (e.g., between strains and stresses) eventually mediated though other measurable and/or internal (i.e., non-measurable) field variables, whose \textit{evolution laws} are part and parcel of the description. The material behavior and, therefore, the CL can be classified either as \textit{path-independent} (also, \textit{history-independent}) if the current state of a point does not depend on its previous states, or \textit{path-dependent}  (also, \textit{history-dependent}) when the current state depends on the history of states experienced by the point. While conservation principles and kinematics are considered axiomatic and epistemic, material modeling is empirical in nature; it is in continuous evolution and it constitutes one of the most important research fields is mechanics.

Since the introduction of Young's modulus in the 19$^{th}$ century \citep{young1807course}, engineers have predominantly defined CLs using the so-called \textit{phenomenological} approach, where experimental observations and physical requirements are distilled into a priori selected analytical ansatz relationships whose parameters are meant to be characteristic of the material \cite{truesdell1965non,holzapfel2000nonlinear,simo2006computational,de2011computational}. Thus, constitutive modeling is traditionally based on data and, due to the limitations of traditional experimental setups, hinged on limited observations on a restricted set of load cases (e.g., inducing uniaxial, biaxial and hydrostatic stress states). On the other hand, CLs were expected to generalize to significantly more complex conditions, a task for which the development of continuum thermodynamics theories and computational techniques has been crucial.

The development of full-field experimental methods such as \textit{digital image correlation} (DIC) \cite{Sutton2013,Sutton2015}, X-ray computed tomography \cite{Withers:2021,Yang:2017,Carrara2016c} and \textit{digital volume correlation} (DVC) \cite{Leclerc2011,Mendoza2019} and advances in the related computational approaches \cite{Pierron2021,Pierron2023} have shifted the constitutive modeling paradigm from a limited- to a large-data regime. In particular, the computational multiscale techniques available to date deliver in-silico observations of arbitrary complex implicit macroscopic behaviors from simple CLs defined at the micro-scale. However, along with the opportunity to obtain better CL comes also the challenge of analyzing the increased amount of data available. In this context, the recent breakthrough in big data analysis and information mining gives a unique opportunity to improve constitutive modeling for applications in mechanics and material science. 
Although the preliminary results obtained in this area bring great promise of efficient and highly accurate predictions, the fast development of this discipline calls for a timely identification of the most promising directions to pursue in order to boost the progress of mechanics and that can introduce a diversity of solution techniques.

\subsection{Classification and nomenclature} 

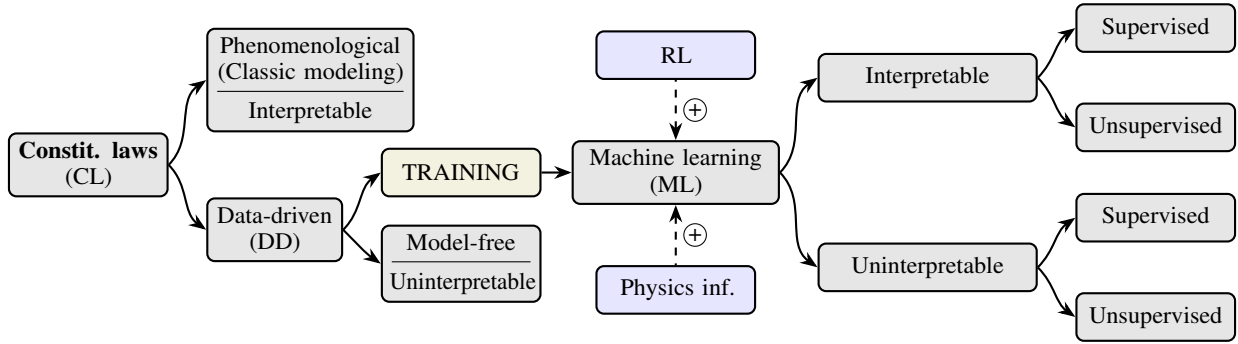
\begin{figure}[hbt!]
\center
	\begin{tikzpicture}
		\footnotesize
\tikzstyle{obj} = [rectangle, draw, thick, fill=gray!20, text width=6em, text centered, rounded corners=1mm, minimum height=2em]
    
\tikzstyle{block2} = [rectangle, draw, fill=yellow!40, text width=4em, text centered, rounded corners, minimum height=2em]
    
\tikzstyle{line} = [draw, -Stealth, thick]

\node[obj](CL){\textbf{Constit. laws} \\ (CL)};
\node[obj, anchor=left, below right = 0cm and .5cm of CL, text width=5em](DD){Data-driven \\ (DD)};
\node[obj, anchor=left, above right = 0cm and .5cm of CL, text width=8em](phen){Phenomenological \\ (Classic modeling) \\  \vspace{0.2em} \begin{tabular} {p{6.5em}} \hline \\ \end{tabular} \vspace*{-6mm} \\  Interpretable};
\node[obj, anchor=left, below right = -.5cm and .5cm of DD](modfree){Model-free \\  \vspace{0.2em} \begin{tabular} {p{4.5em}} \hline \\ \end{tabular} \vspace*{-6mm} \\ Uninterpretable };
\node[obj, anchor=left, above right = 0cm and .5cm of DD, fill=olive!10](train){TRAINING};
\node[obj, anchor=left, right = 0cm and .4cm of train, text width=8em](ML){Machine learning\\ (ML)};
\node[obj, anchor=right, above right = 0.5cm and .5cm of ML, text width=8.5em](int){Interpretable};
\node[obj, anchor=right, below right = 0.5cm and .5cm of ML, text width=8.5em](uint){Uninterpretable};
\node[obj, anchor=right, above right = 0cm and .5cm of int](int_sup){Supervised};
\node[obj, anchor=right, below right = 0cm and .5cm of int](int_usup){Unsupervised};
\node[obj, anchor=right, above right = 0cm and .5cm of uint](uint_sup){Supervised};
\node[obj, anchor=right, below right = 0cm and .5cm of uint](uint_usup){Unsupervised};
\node[obj, anchor=right, below= 0.8cm of ML, fill=blue!10](PI){Physics inf.};
\node[obj, anchor=right, above= 0.8cm of ML, fill=blue!10](RL){RL};

\path[-Stealth,thick]
     (CL.east)  edge [out=-45, in=145] (DD.west)
     (CL.east)  edge [out=45, in=-145] (phen.west)
     (DD.east)  edge [out=-45, in=145] (modfree.west)
     (DD.east)  edge [out=45, in=-145] (train.west)
     (train.east)  edge (ML.west)
     (ML.east)  edge [out=-45, in=145] (uint.west)
     (ML.east)  edge [out=45, in=-145] (int.west)
     (int.east)  edge [out=-45, in=145] (int_usup.west)
     (int.east)  edge [out=45, in=-145] (int_sup.west)
     (uint.east)  edge [out=-45, in=145] (uint_usup.west)
     (uint.east)  edge [out=45, in=-145] (uint_sup.west);
     
  \path[Stealth-,thick,dashed]   
     (ML.north)  edge node[circle, inner sep=0pt, minimum size=3mm, draw,right=1mm,solid,thin]{+} (RL.south) [midway, label=right:Plane]
     (ML.south)  edge node[circle, inner sep=0pt, minimum size=3mm, draw,right=1mm,solid,thin]{+} (PI.north);

	\end{tikzpicture}
\caption{Characteristic features of the approaches in the constitutive modeling context.}
\label{fig:char}
\end{figure}

Available techniques that deal with large data 
are often referred to as \textit{data-driven} (DD) approaches, due to their strict dependence upon a set of experimental or numerical observations of the material behavior
(Figure~\ref{fig:char}). In general, each method deploys a set of algorithms, assumptions and procedures, whose purpose is to analyze the available data and deduct useful information to describe the behavior of a certain material or class of materials. 
Since the recent breakthroughs in deep learning, machine learning (ML) techniques have received renewed interest in applications 
in mechanics and materials. Particularly the rise of sparse identification and discovery \cite{schmidt_distilling_2009,brunton2016discovering}, Physics Informed neural networks (PINNs) and operator learning approaches \cite{lagaris1998artificial,raissi2019physics} for the solution of forward and inverse problems involving partial differential equations (PDEs) \cite{liu_neural_2020,lu2021learning} have had a significant impact on the computational engineering field, shifting the attention to combining data with physics. The great promise of fast and highly accurate predictions for mechanistic simulations as well as cost reduction in an industrial setting 
has led to numerous attempts to integrate a wide spectrum of these techniques in the simulation workflow and also to facilitate material innovation. Combining physical constraints and data in the context of ML approaches has shown potential towards moderating the need for large training datasets \cite{fuhg2022physics}.

For the purpose of this review on DD constitutive modeling in solid mechanics, with the intent of providing a structured presentation, we propose for the available DD approaches the classification outlined in Figure \ref{fig:char}. We distinguish among two broad categories with corresponding subcategories, as follows:
\begin{itemize}
    \item methods based on \textit{machine learning} (ML) techniques. Within this category, we distinguish among  
    \begin{itemize}
    \item \textit{uninterpretable} (black-box or grey-box) approaches. This subcategory includes methods that obtain CLs in which the relation existing between inputs and outputs cannot be physically explained. These methods are also sometimes referred to as \textit{encoding} or \textit{learning} of CLs. Prominent ML techniques used in uninterpretable approaches are those based on many different types of neural networks (NNs) (see Section \ref{ML_approaches}). The difference between black-box and grey-box approaches is that the latter encodes some known information about the physical system in the learning framework; this augments the clarity of the learned model but still does not enable interpretability. Recently, sparsification approaches for NN models aim to balance expressivity and interpretability.
    \item \textit{interpretable} approaches. Within this category fall the techniques that aim at defining an \textit{interpretable} analytical expression for the CL, namely a relation between input and output quantities (e.g., strains and stresses) through mathematical operators and parameters whose role can be physically explained. In the literature, some of these methods are referred to as methods for \textit{automated discovery} of CLs from data, whereby the term \textit{discovery} is used to refer to simultaneous \textit{model selection} and \textit{parameter identification}. The special case of parameter identification, in which the functional form of the model is known a priori and only the unknown parameters are identified, collapses in terms of scope with the parameter identification of traditional constitutive modeling, but is now sometimes also carried out with ML. Prominent ML techniques used in interpretable approaches are symbolic regression and sparse regression (see Section \ref{ML_approaches}).
    \end{itemize}
    In ML-based approaches, the definition of the operators and parameters takes the name of \textit{training} or \textit{learning} process, whereas their structure and configuration is also known as \textit{architecture};
 \item \textit{model-free} approaches. These aim at integrating the material observations into the solution stream of a solid mechanics problem. In other words, they identify the solution of the system of PDEs governing a solid mechanics problem within the collected material observation pairs, thus bypassing an explicit analytical link between them. Note that, while ML-based methods deliver a CL mapping the input to the output which, once obtained, does not need any interaction with the original data, model-free methods cannot be set apart from the dataset they try to represent, since the dataset becomes part of the solution.
\end{itemize}


The above classification places a strong emphasis on interpretability. Interpretable approaches 
often lead to parsimonious (i.e. simple) CL representations, and parsimony is known to counteract the issue of overfitting which comes as a side effect of over-parametrization and limited training data. Additionally, parsimonious representations can enable generalization capabilities beyond the range of input data, given that the physics remains consistent (follows the same discovered law) out of that range.

An important aspect not contained in the above classification is the distinction between \textit{supervised} and \textit{unsupervised} methods. In ML, training is performed in a \textit{supervised} fashion when matching pairs of labeled input and output data are used, or in an \textit{unsupervised} manner if no labeled output data are available or in the absence of a bijective mapping between input and output quantities. 
Cases where the learning process objective is to minimize the deviations between the model predictions and the labeled data are part of the first category, while the methods whose goal is to detect hidden patterns or relationships within the data fall in the second. 
In the context of constitutive modeling, input and output quantities are, in the simplest case, strains and stresses (the situation becomes more complex for path-dependent material models, see Section \ref{sec::intro}). For this reason, we denote as supervised the methods that require training data in the form of stress-strain pairs, and as unsupervised those that do not require such pairs. The distinction is important because, while strains can be measured (almost) directly, stresses can only be computed from measured forces under very simple loading conditions, such as uniaxial tension, and are therefore not available in general cases. Unsupervised methods formulate the learning procedure in such as way as not to rely on the availability of stress-strain pairs; to compensate for the lack of such pairs, they typically rely on the enforcement of physics constraints which can be formulated in terms of realistically measurable data, such as displacements and forces.

Physics constraints are not exclusive of unsupervised methods; approaches of all types can augment learning by imposing the fulfillment of such constraints (e.g., global or local equilibrium, governing PDEs), in which case they are sometimes also denoted as \textit{physics informed}  (Figure~\ref{fig:char}).
To enforce these constraints, mainly three procedures can be deployed, either alone or in synergy: i.) the selection of an architecture that a priori satisfies them, ii.) penalizing the selection of parameters leading to the violation of the constraints, iii.), the a posteriori rejection of optimized models which are incompatible with the constraints following predefined rejection/acceptance policies. Another ML paradigm is the so-called \textit{reinforcement learning} (RL), whose aim is not limited to the optimization of a predefined architecture but extends to learning a set of rules that allows for the improvement of the architecture itself by maximizing its predictive capability following a given metric  (Figure~\ref{fig:char}). This is normally done in a trial-and-error fashion by casting architectures whose performances are tested against the available data; a reward or a penalty is then given following a set of predefined policies. Note that the described methodologies are not mutually exclusive, but can be coupled in order to achieve a better description of the material behavior.
 The main characteristic features of the approaches to define a CL are summarized in Figure~\ref{fig:char}, while their detailed description will be provided later. 

\subsection{Objective and organization of the paper}
The primary goal of this review is not to provide an exhaustive list or a ranking of DD attempts in the field of solid mechanics but, rather, a map of how DD techniques can be used to advance the constitutive modeling field. We provide an overview of how DD approaches can exploit the large amount of available observations in order to obtain richer CLs able to accurately predict the behavior of different materials. The available literature is reviewed and classified following the main aspects summarized in Figure~\ref{fig:char} and whether the behavior they try to represent (e.g., elasticity, plasticity, damage) is path-independent or path-dependent. Whenever possible and relevant, we also overview the mechanics/physics knowledge that the various approaches embed, the capability to generalize to situations different than those represented by the data they are exposed to, and the amount of data they need to provide reliable results (namely, the \textit{data hunger}). In the overview, we attempt to pinpoint the specific advantages and drawbacks of some of the current state-of-the-art techniques and our perspectives on the future of DD constitutive modeling for solid mechanics problems. Although not treated extensively, the strict relation between modern experimental techniques and DD approaches is also hinted at, especially regarding the need for integrated frameworks where the experiments are designed to support the definition of accurate and realistic CL. 


The paper is structured as follows: in Section 2 we summarize some earlier reviews in the context of DD approaches in solid mechanics and we provide a short overview of DD methods to establish a common ground for the remainder of the paper. Section 3 focuses on data sampling approaches, a central aspect of every approach that focuses on data. Section 4 reviews DD CLs for path-independent problems with a focus on small strain elasticity and finite strain hyperelasticity. Section 5 reviews DD CLs for a range of path-dependent problems, from plasticity, viscoelasticity, damage, fracture, and fatigue, to problems in multiphysics. Section 6 presents some thoughts on guidelines for validation and verification and proposes a set of possible performance metrics, while Sections 7 and 8 include respectively a discussion on the current limitations of the reviewed approaches and the main conclusions.

\section{DD methods outline and earlier reviews}

\subsection{Overview of the available DD methods}

In the following, a non-exhaustive list of important DD methods is provided, as they are used either alone or in combination for the different constitutive models discussed in the following sections. For most of these models, different variations exist that allow for either classification of data or fitting regression curves. Since in constitutive modeling, we are usually concerned with the prediction of quantities instead of labels, the following summary mainly focuses on regression tasks.

\subsubsection{ML approaches}
\label{ML_approaches}

\textbf{Interpretable approaches.} Symbolic regression and sparse regression are probably the closest ML techniques to traditional constitutive modeling (based on parameter identification of assumed laws) and have the advantage that functional dependencies between inputs and outputs are highly interpretable, and constitutive modeling constraints can be incorporated, which typically leads to high extrapolation power.
Both approaches are based on finding an analytical function that best fits a given dataset. In general, approaches that derive mathematical expressions from targeted experimental data are commonly referred to as symbolic regression methods; hence, also sparse regression can be interpreted as a specific instance of symbolic regression (see also the extensive review in \cite{Kissas2024}, where symbolic regression methods, including sparse regression, are viewed through the lens of operations on graphs). However, in the literature very often symbolic regression is used to denote methods based on genetic algorithms. In what follows, we adopt this terminology; hence, we refer to methods that are based on genetic algorithms as symbolic regression and treat sparse regression separately.

In symbolic regression, building blocks such as mathematical operators and constants are combined with commonly applied analytical functions such as sinus or logarithm using genetic algorithms \cite{koza_genetic_1994,billard2002symbolic}. In the physical sciences, symbolic regression was popularized by \cite{schmidt_distilling_2009}, and an overview with applications to material science is provided in \cite{wang_symbolic_2019}.
Different software packages can be found in the literature, e.g., GPTIPS \cite{searson_gptipsopen_2010}, Eureqa \cite{dubcakova_eureqa_2011}, gplearn \cite{stephens2016genetic}, AIFeynman \cite{udrescu2020ai}.
A recent comparison between common open-source implementations can be found in \cite{la2021contemporary}.
The disadvantage of these methods is the long training time of the genetic algorithms without guaranteed convergence to a good solution.
Furthermore, in comparison to other methods, which are more automatized, a high degree of user knowledge is required since the space of mathematical expressions has to be a priori provided by the user. Additionally, due to the commonly low number of trainable parameters, the method is not as expressive and might struggle with highly complex functional dependencies.

In contrast to symbolic regression, sparse regression \cite{santosa_linear_1986,frank1993statistical,tibshirani1996regression} discovers a symbolic model expression from a predefined catalog (often denoted as a library) of candidate models. The strength of sparse regression is its computational efficiency and, in the frequent case of a convex objective function, guaranteed convergence. The idea of using sparse regression in the physical sciences was initiated by \cite{brunton2016discovering}. Recently, sparse regression has gained increasing popularity for material model discovery \cite{flaschel_unsupervised_2021,flaschel_automated_2023,wang_inference_2021}. Also, in this case, user knowledge is required since the catalog of material models is chosen upfront by the user. However, recent research is trying to automatize this task towards a self-generated library \cite{Kissas2024}.

\textbf{Uninterpretable approaches.} 
In recent years, artificial NNs and especially deep NNs have been widely used for many complex regression tasks and NNs are probably the most encountered type of ML method in the context of DD constitutive modeling. 
Their historically first and simplest form, known as feed-forward neural NN or fully connected NN, takes vector-valued inputs and maps them to the vector-valued output using only forward-directed connections between the hidden layers of the network, that consist of several hidden nodes. 
Other forms include sequential modeling architectures such as recurrent (or sequential) NNs (RNNs) \cite{sherstinsky2020fundamentals}, Gated Recurrent Unit (GRU) \cite{nosouhian2021review}, Long Short-Term Memory (LSTM) \cite{sherstinsky2020fundamentals}, and Transformers \cite{vaswani2017attention}, which are based on internal states that allow for processing time- and history-dependent datasets. Lastly,
convolutional NNs (CNNs) \cite{alzubaidi2021review}, attention networks \cite{wang2018non}, and graph NNs (GNNs)\cite{zhou2020graph} are used to analyze visual imagery and graph dependencies, respectively.
For more information on these types of NNs, we refer to \cite{goodfellow2016deep}.
One important reason for the popularity of NNs is that all the major current software implementations are based on Automatic Differentiation \citep{baydin2018automatic}.
The most commonly employed open-source libraries are
Tensorflow \citep{abadi2016tensorflow},
Pytorch \citep{paszke2019pytorch} and
JAX \citep{frostig2018compiling}.
Other than their flexibility, the advantages of NNs include their ease of usage and their ability to deal with very large datasets. An important positive aspect of NNs is that they allow to weakly incorporate constitutive model constraints. 
On the other hand, even though NNs with arbitrary widths or depths are theoretically able to approximate any function \citep{hornik1989multilayer,lu2017expressive}, from an engineering point of view with naturally limited computational resources, convergence cannot be guaranteed.
Furthermore, NNs do not offer exact inference properties, their analytical function is hard to interpret due to the number of trainable weights, and they are reliant on a large amount of user-chosen parameters (hyperparameters). Also, performing Bayesian inference using NNs, known as Bayesian NNs (BNNs)\cite{lampinen2001bayesian}, is popular in the ML community for uncertainty quantification (UQ). Here, additional constraints in constitutive modeling can be enforced in various ways \cite{swiler2020survey}. However, the process typically requires a higher understanding of statistical learning theory in comparison to enforcing constraints with NNs.
\par
NNs and Bayesian NNs are widely used for many computationally expensive scientific problems for performing tasks such as inverse modeling or UQ. However, these models are confined to the mapping between a predefined set of input and output data. They cannot be directly applied to control or experimental design problems due to the difficulty in learning optimal decision-making strategies in dynamic and uncertain environments. To address this issue, an RL technique that hinges on the concept of dynamic programming is employed \cite{arulkumaran2017deep}. RL is a type of ML model in which the agent or an NN model learns the best actions to make decisions by interacting with the environment. The decision-making process in RL is executed by defining states, actions, and rewards. During training, the data from the environment known as the state is provided to the agent to assess the reward.  The main objective of RL is to enhance the model's policy or value function, enabling it to take the best actions over the iterations known as episodes. There are mainly two types of RL models, namely, model-free and model-based RL. Model-free RL algorithms, such as Q-learning \cite{clifton2020q} and Policy Gradient algorithms \cite{grondman2012survey}, focus on learning the optimal solution by directly interacting with the environment. On the other hand, model-based RL algorithms such as Dyna \cite{sutton1991dyna} and AlphaGo Zero \cite{silver2018general} depend on planning based on the learned model.  

Support vector regression (SVR) is well established in the ML community for real-valued function estimation \citep{hearst1998support}. Rooted in statistical learning theory, SVR is based on its well-known classification-based counterpart known as support vector machines (SVM), which try to systematically find a linear hyperplane able to cluster input and related output data in classes. However, the latter assumption fails to represent the classification when the mapping between input and output parameters is non-linear. In such cases, the 'kernel trick' \citep{hofmann2006support} is employed to convert the original data space into a higher dimensional one where a linear hyperplane is again a suitable ansatz. Considering the domain and codomain of an unknown function composed of classes representing a restricted range of values, it is possible to endow the SVM with regression capabilities leading thus to the SVR. To avoid overfitting, SVR penalizes predictions farther away than a specific value from the desired output in a convex loss function. For constitutive modeling in solid mechanics, SVR has traditionally been far less commonly employed compared to other ML methods although different libraries offer the possibility to use SVR. The main open-source software library used in the literature is libSVM \cite{chang2011libsvm}. Alternatively, the Scikit-learn \cite{scikit-learn} also has an implementation of SVR, which is, however, also based on libSVM.
The advantages of SVR methods are that they are robust against outliers in the data, are easy to implement, their computational complexity does not depend on
the dimensionality of the input space, and, when well fitted, they tend to have good generalization capabilities \citep{smola2004tutorial,awad2015support}. Furthermore, due to the convexity of the loss function, SVR training delivers a unique solution that makes training in comparison to other advanced ML methods easier. Constraints generally encountered in constitutive modeling can be enforced by adjusting the loss function \citep{lauer2008incorporating}. However, the method has problems in the big data domain and with datasets that contain a significant amount of noise. Furthermore, in comparison to GPR, there is no probabilistic explanation for the fitting and no exact inference.

\par 
Apart from the above-mentioned models for regression tasks, feature extraction models and probabilistic generative models are other popular ML areas that are widely used in the scientific domain. Feature extraction is used to reduce datasets into their informative and non-redundant parts. For more details, we refer to \cite{guyon2008feature}. They usually involve some form of dimensionality reduction, and common techniques of this kind are principal component analysis (PCA) \cite{jolliffe2016principal}, autoencoders \cite{bank2023autoencoders}, and clustering techniques such as k-means clustering. Other popular ML domains are probabilistic generative models such as restricted Boltzmann machine (RBM) \cite{zhang2018overview}, variational autoencoder (VAE) \cite{kingma2013auto}, generative adversarial network (GAN) \cite{goodfellow2020generative}, flow-based generative models (or Normalizing Flows) \cite{dinh2016density} or diffusion models (DM) \cite{croitoru2023diffusion}, which have been successfully implemented for various tasks such as dimensionality reduction, inverse surrogate modeling \cite{padmanabha2021solving}, image generation \cite{goodfellow2020generative}, anomaly detection \cite{di2019survey}, and many others. Due to the relatively small datasets generally encountered in constitutive modeling, these methods have not yet seen a lot of interest in the community. However, with the ever-increasing availability of data, they may soon become more important. 

\textbf{Bayesian inference and statistical learning.} Another important class of ML methods is models based on Bayesian inference. They revolve around using Bayes' theorem to find the probability distribution of parameters or functions given observed data and prior distributions. Since often no closed-form solutions for the posterior distributions exist, efficient simulation algorithms like Markov Chain Monte Carlo (MCMC), Metropolis-Hastings (MH), and Hamiltonian (or hybrid) Monte Carlo (HMC) are applied to find approximations.
For more information, we refer to \cite{korb2010bayesian}. 
Analytical posterior distributions can be found by using Gaussian processes as priors. The resulting interpolation method is typically known as Gaussian process regression (GPR) or Kriging \cite{rasmussen2003gaussian}. 
In the context of ML CL modeling, GPR is the most encountered form of Bayesian regression due to its simplicity. Different open-source software packages for GPR are available, e.g., DACE \cite{lophaven2002dace}, DiceKriging \cite{roustant2012dicekriging} or GPyTorch \cite{gardner2018gpytorch}. GPR methods have significant advantages compared to other ML models: they have rigorous convergence guarantees, yield exact inference in the absence of noise, allow for probabilistic error estimation, and are fully trainable with only a limited amount of user-chosen parameters \citep{rasmussen2003gaussian}.
The major issue of GPR is that it scales cubically with the size of the dataset; this has historically prevented it from being used in the big data context \citep{liu2020gaussian}. Overall, utilizing GPR for constitutive modeling has been a relatively recent phenomenon in comparison to other ML models.
Finally, classical but proven techniques of statistical learning theory should not be neglected when deciding which predictive tool to use. Methods like k-nearest neighbor, random forests, or spline interpolation might, depending on the complexity of the data, even be the best-performing methods for certain tasks. For a textbook on these traditional techniques, we refer to \cite{james2013introduction}.

\textbf{Closing remarks.} As pointed out, all of the presented methods have their own strengths and weaknesses, and similar to most application areas in engineering there is also no single best method to apply for constitutive modeling. The choice of which method to use is typically dependent on the size of the dataset, the type of the data source, the required level of interpretability, the degree of dimensionality of the inputs and outputs, and the complexity and amount of noise in the data.
Moreover, in the end, the choice of an ML method usually boils down to user knowledge of specific methods and algorithms and the availability of codes; facilitating this process could be beneficial for the broader community. In this regard, due to the number of historical papers and available tutorials, NNs can certainly be seen as the easiest and most flexible method for inexperienced users. Approaches that initially show a steeper learning curve might however have some significant advantages for specific applications and goals.

\subsubsection{Model-free approaches}
 As mentioned earlier, model-free approaches first proposed by \cite{kirchdoerfer2016data} do not explicitly create material models or their surrogates to be integrated into computations but, instead, directly inform the forward problem with a set of discrete material behavior observations (the so-called \textit{material data set}). The main idea is to relax the requirement of the classical solution procedure of a mechanical boundary value problem that seeks the solution as the intersection between a set of basic governing equations (e.g. equilibrium, compatibility) and the CL. Instead, the model-free approach identifies for each point of the domain the state within the available material observations that is closest to the subset of points satisfying compatibility and equilibrium. Hence, the model-free solution strategy relies on the definition of a discrete quantity, representing a distance induced by a given metric in the state space, which attains its minimum in correspondence with the material data point that best represents the solution under the given boundary conditions \cite{Carrara2022}. The aim of this approach is to minimize assumptions on the modeling part by relying only on discrete observations of the material behavior. This makes the model-free approach data-hungry \cite{he2020physics, bahmani2022manifold} since it is unable to generalize the observed behavior. For the same reason, it poses limitations to the description of dissipative CLs, where the major challenge is to ensure the representability of the material behavior without the introduction of a priori-defined dependencies from postulated internal/history variables or oversized datasets.

\subsection{Earlier reviews}



There are several recent reviews in the context of DD approaches for a wide array of solid mechanics problems. We summarize here the main contributions towards three main fields: i. single or multiscale approaches to represent the material behavior, ii. design of new (meta-)materials with prescribed behavior (namely, material behavior optimization) and, iii., synergistic approaches integrating experimental mechanics and ML methods.

\textbf{Material behavior representation.} In \cite{montans2019data} a review of DD modeling approaches in engineering is presented that aims to introduce the reader to a variety of approaches and applications. With a focus on \textit{virtual twins}, namely computational frameworks replicating the material behavior that support and extend the experimental practice, \cite{chinesta2020virtual} discusses efficient approaches for the exploitation of data in computations beyond classical parameters calibration. More focused on multiscale approaches, surrogate models, and design optimization of composite materials and structures, \cite{liu2021machine} revises the available approaches employing NNs, while \cite{peng2021multiscale} focuses more on physics-informed ML and tries to identify applications and opportunities in the general context of computational mechanics. Also, \cite{aldakheel_what_2022} extends the review to the whole ecosystem of ML approaches and tries to outline the future perspectives of the field. Focusing on mechanical properties, \cite{guo_artificial_2021} discusses the importance of data collection, generation, and preprocessing towards applications in multiscale materials design.  

\textbf{Design of meta-materials.} \cite{luca_reviewing_2018} reviews the advancements driven by ML tools towards material design innovation, while in \cite{bock2019review} a classification among descriptive, predictive, or prescriptive is utilized to map ML approaches to problems of parameter calibration, material characterization, and material design and optimization.
Focusing on material discovery and the materials genome, \cite{suh2020evolving} discusses how ML approaches can take advantage of mechanical and chemical datasets to propel the design of new meta-materials with improved properties, while \cite{huang2020artificial}  and \cite{morgan2020opportunities} review respectively the methods available and the challenges and opportunities involved. 

\textbf{Synergy between experimental and DD methods.} Various hierarchical identification procedures and related data-reduction methods to accelerate the exploitation of the experimental information are accounted for in \cite{neggers2018big}, while \cite{brodnik2023perspective} reviews the ML approaches involving data from acoustic emission and resonant ultrasound spectroscopy and points out that including known physical and mechanistic relationships in the ML approaches increases the reliability of the trained models. The integration of experimental and ML methods is exploited extensively in the area of biomechanics. With this focus, \cite{tepole2022data} reviews aspects of the solution of biomechanics problems using DD methods based on patient-specific data and highlights how this enables an automatic consideration of the case-wise variability of the material parameters and of the uncertainty propagation. 

Despite the number of review articles published, to the best of the authors' knowledge a review that solely targets constitutive modeling approaches in solid mechanics is still lacking. With this paper, we aim to fill this gap.

\section{Data sampling and design of experiments}\label{sec::DataSampling}
One of the open problems in DD material modeling is the dependence of the performance on the amount of available data. One possible remedy to the unavailability of sufficient data is to train the ML tools on less but qualitatively more relevant data, i.e. data that capture the major complexities of the mapping. This can be achieved by relying on pertinent data sampling strategies. In this section, we briefly review some of these strategies that have been employed in the literature for DD material modeling. We differentiate between one-shot and sequential sampling approaches.

The design of sampling strategies can be paralleled to experimental design. Experimental design denotes the systematic planning and structuring of lab experiments (e.g. loading paths, types of experiments) to gather valuable and relevant data from lab tests to calibrate constitutive models. While both sampling strategies and experimental design aim to efficiently extract the most meaningful data, experimental design brings an additional layer of complexity. It must not only consider the theoretical significance of the data but also the feasibility and efficiency of conducting the experiment in a real-world lab setting. Practical limitations, such as equipment availability, time constraints, and material properties, often dictate what can be accomplished in the lab.
After discussing one-shot and sequential sampling approaches, we briefly overview the works that utilize deep RL for experimental design, showcasing its potential in optimizing and informing experimental setups.

\begin{figure}[h!]
\begin{subfigure}[b]{0.33\linewidth}
        \centering
    \includegraphics[scale=0.28]{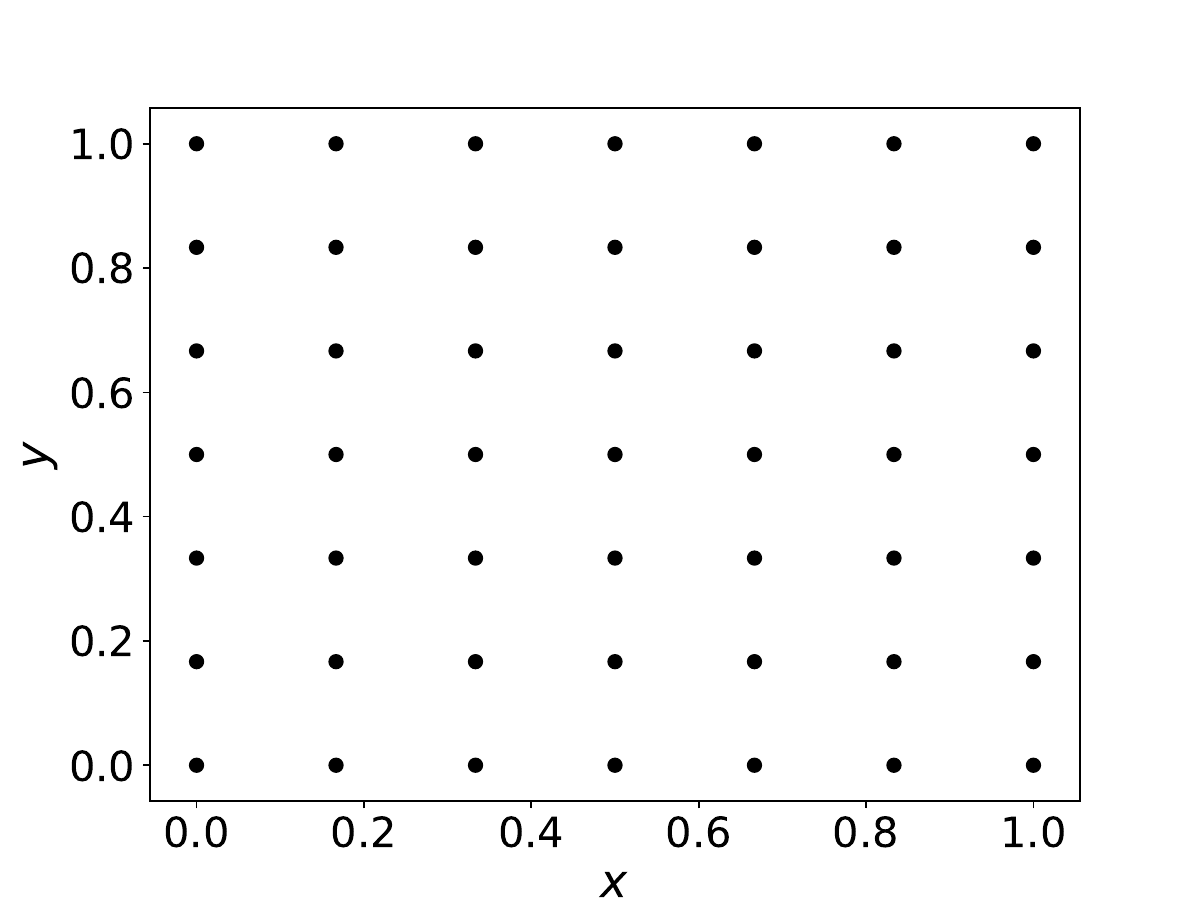}
    \subcaption{Grid sampling}\label{fig::gridSampling}
\end{subfigure}
\begin{subfigure}[b]{0.33\linewidth}
        \centering
    \includegraphics[scale=0.28]{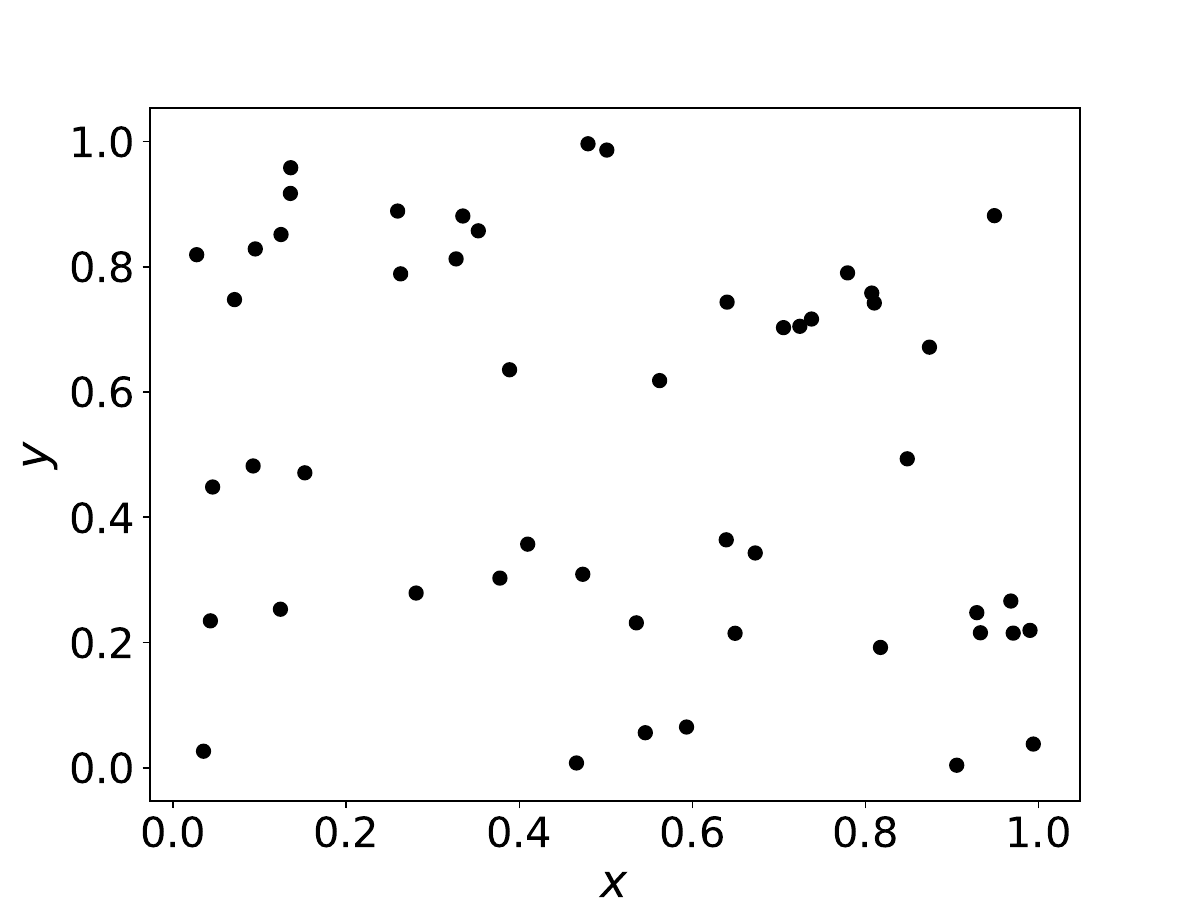}
    \subcaption{Uniform sampling}\label{fig::uniSampling}
\end{subfigure}
\begin{subfigure}[b]{0.33\linewidth}
        \centering
    \includegraphics[scale=0.28]{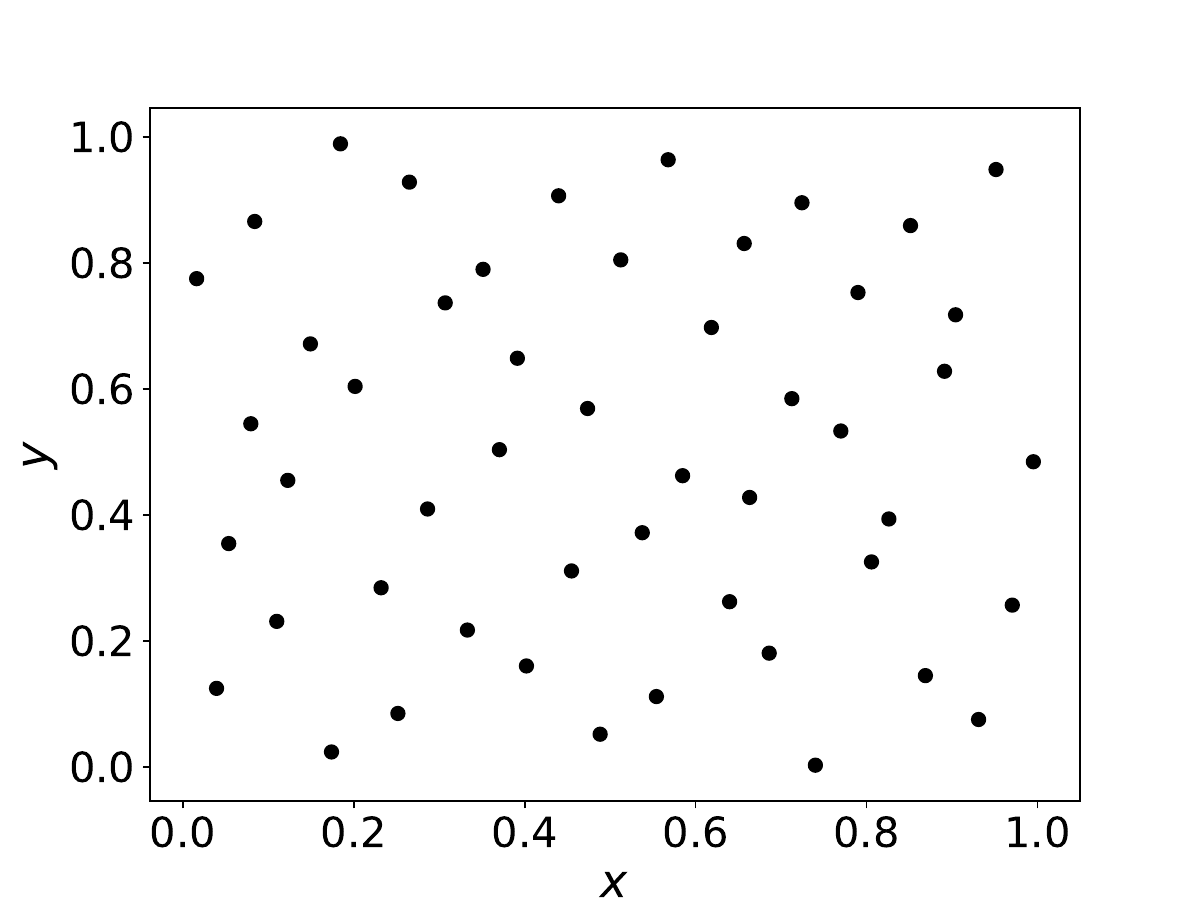}
    \caption{Latin hypercube sampling}\label{fig::lhsSampling}
\end{subfigure}
    \caption{Different one-shot sampling strategies. (a) Grid sampling. The input domain is discretized using equidistant samples. (b) Uniform sampling. The sample points are generated from a uniform distribution in the input domain. (c) Latin hypercube sampling. The points are sampled such that there is only one point at each section of a divided input space. }
\end{figure}

\paragraph{One-shot sampling.}
One-shot sampling is characterized by the determination of the sample size and points in a single stage. To this end, the input domain for both time-dependent and time-independent inputs has to be known.
Non-temporal inputs are characterized by a fixed range of interest, i.e. a sampling interval, or a probability distribution function. Since sampling from a probability distribution is straightforward, in the following we focus on deterministic input domains.
The simplest way of generating data for non-temporal inputs is
grid-based sampling (Figure \ref{fig::gridSampling}) where a few equally spaced values are selected for each parameter. Apart from its simplicity, this type of strategy has several other advantages, including easy setup for parameter sensitivity studies and numerical integration, and of course its space-filling properties.
Grid-based sampling strategies have been applied in a variety of publications \cite{qu2021towards,vlassis2021sobolev,FUHG2021103522}.

A major limitation of grid-based sampling is their
"collapsing" property \cite{crombecq2011efficient} which essentially means that sample points may have the same coordinate value when projected onto a parameter axis. 
This has the undesirable effect that, when one of the design parameters has (almost) no influence on the mapping, then two points whose only difference is this property can essentially be seen as the same point. This means that for the purposes of generating a surrogate for the mapping the same point will be evaluated twice. 
Hence, the two crucial criteria for one-shot sampling designs are space-filling and non-collapsing properties.
An obvious way to avoid a collapsing design would be to use uniform random sampling. However, especially if a small quantity of sample points is involved, random sampling approaches tend to be not reliably space-filling, i.e. they show clustering behavior and fail to provide points in large portions of the region, as seen in Figure \ref{fig::uniSampling}.
For this reason quasi-random techniques such as Latin hypercube sampling \cite{stein1987large} are often preferred (Figure \ref{fig::lhsSampling}). This method aims to generate a space-filling design that is non-collapsing. This strategy is employed in various works in the context of DD material modeling
\cite{lu2019data,feng2022finite,FUHG2022105022}.
In some applications, due to indirect access to the input space (i.e. the input domain cannot be explicitly sampled), different strategies have to be employed.
E.g., \cite{fuhg2022physics}  proposes a sampling approach that generates space-filling points in the invariant space corresponding to a bounded domain of the deformation gradient tensor. The approach builds a space-filling sampling strategy based on simulated annealing, which provides more efficient and reliable physics-informed constitutive models than quasi-random sampling in the deformation gradient space.

Sampling from temporal inputs is more complicated as now a sequence of points has to be generated. This is e.g. required when the response depends on a specific three-dimensional loading path.  
Some aspects of effective temporal sampling are still an open issue in the literature, e.g. how to sample temporal curves in a space-filling manner.
Different approaches have been proposed.
In strain-controlled loading, \cite{mozaffar2019deep} defines an upper and lower bound for each individual strain component and a fixed number of time steps. The authors then propose to generate equally spaced points in the chosen bounds over the time frame using random sampling. These points are used as control points of an interpolator in order to generate smooth loading curves. Similar approaches are used in \cite{gorji2020potential,abueidda2021deep, bonatti2022cp}.
\cite{wu2020recurrent} samples the loading path as a random walk which changes its direction and step size using realizations of independent uniform random variables. The same method is employed in \cite{bonatti2022cp}.
Other approaches \cite{logarzo2021smart} define the load paths directly as realizations of a multivariate Gaussian process.

Even though these methods have proven to be successful, overall, reliable and goal-oriented sampling of sequentially-based material responses still remains an open issue. Since non-temporal data-generation techniques are in general more reliable, DD modeling frameworks that transform a many-to-one mapping (as seen in plasticity) into one-to-one mappings can be considered more reliable.
\par
\textbf{Sequential/adaptive sampling.}
The problem with one-shot sampling is that, without a target function, it is difficult to predetermine an optimal or appropriate sample size as well as the ideal sample placement. This is especially true for temporal input parameters.
For these reasons sequential or adaptive sampling techniques have been proposed in the broader surrogate modeling community \cite{settles2009active,liu2018survey,fuhg2020state}. These techniques have been only sparingly applied in the context of DD material modeling.
However, some approaches have been proposed in this field that could pave the way forward.
To the best of the authors' knowledge,
\cite{knap2008adaptive} is the first contribution that employs an adaptive sampling scheme for multiscale problems. The authors use multivariate kriging as a fit for a path-dependent DD material model. The proposed method is based on using the variance estimation of the kriging interpolant as an error estimate in the local material point routine of a finite element model. In essence, the surrogate takes the stress, pressure, temperature, and hardening variable of the current load step as input and returns the approximate values of the lower scale model, such as the directionality of the plastic velocity gradient, as well as an error estimate. If this error estimate is larger than a tolerance value, then a new data point is added at this input point and the surrogate model is retrained. This framework is refined in \cite{leiter2018accelerated}.
Similar approaches are proposed in 
\cite{rocha2021fly, kalina2022fe}, but instead of relying on the error estimate of the surrogate, the model is adaptively refined when convergence in the local integration loop at a material point of the macroscopic structural problem cannot be achieved. 
In the context of model-free plasticity modeling \cite{karapiperis2021data} proposes a related idea.
Overall, especially concerning path- and history-dependent material modeling, adaptive sampling strategies are largely underused. They have the potential to offer a reliable framework to train more robust DD material models.
\par
\textbf{Deep RL for experimental design.}
Deep RL is a subdomain of RL that combines the concepts of RL and deep NNs for 
tackling complex tasks involving continuous state and action spaces in an uncertain and dynamic environment. In the ML community, there are various deep RL models, such as Deep Q-Network (DQN) \cite{roderick2017implementing}, Deep Deterministic Policy Gradient (DDPG) \cite{tiong2020deep}, and others that have been used for a wide range of problems. These include learning acquisition functions for Bayesian optimization \cite{volpp2019meta}, trading and finance \cite{zhang2020deep}, robotics manipulation \cite{kalashnikov2018qt}, controls \cite{duan2016benchmarking}, experimental design \cite{villarreal2023design}, and many more. Recently, \cite{wang2021non} developed a meta-modeling game framework based on RL, where two ML agents compete against each other to generate experimental data for calibrating a constitutive model. Out of all possible design options for laboratory experiments, the developed deep RL framework can determine sophisticated experimental designs that capture the relevant information to train reliable constitutive models. With a focus on using deep RL for expensive experimental data, \cite{villarreal2023design} develops a framework that combines deep RL and Kalman filters for calibrating materials models. 



\section{DD modeling for path-independent CLs}\label{sct:pathindep}

This section reviews the main DD modeling approaches proposed for path-independent CLs, namely for conservative and thus fully reversible material behavior. A key feature of this type of materials is that the current state of a point is uniquely defined by the current value of the strain in that point. In the following, starting with linear elasticity and continuing with the finite-strain case, we review the main related contributions in the context of DD modeling. 


\subsection{Small-strain elasticity}

The simplest assumptions in phenomenological constitutive modeling of solid materials are the assumption of small strains and the often accompanying assumption that the stress state depends linearly on the strain state $\boldsymbol\sigma = \mathbb{C}\colon\boldsymbol\varepsilon$, where $\boldsymbol\sigma$, $\boldsymbol\varepsilon$ and $\mathbb{C}$ denote the Cauchy stress tensor, the infinitesimal strain tensor and the elasticity tensor, respectively.
Under these assumptions, the material characterization objective boils down to identifying the parameters that govern the linear stress-strain relation, i.e., the components of the elasticity tensor $\mathbb{C}$.
Although not denoted as such, this parameter identification problem has often been faced from a DD perspective since its inception. Despite the simplicity of the problem, ML opens new doors for elastic property prediction. 

While the response of homogeneous materials appears straightforward to model, mesoscopically heterogeneous materials such as multiphase materials and composites exhibit more complex stress-strain (or energy-strain) relations at the macroscopic scale. Hierarchical modeling is a powerful tool to describe the macroscopic material responses of such materials; the idea is to conduct lower-scale simulations of representative volume elements (RVEs), which capture the characteristic topology of the heterogeneous material, to predict the effective material behavior at the macroscopic scale. Prerequisites to such lower-scale simulations are, first, that the material properties of the different phases at the micro- or meso-scale are known and, second, extensive computational effort, especially in the three-dimensional case. Hierarchical methods such as FE$^2$ require the execution of a lower-scale simulation whenever the unknown stress state needs to be calculated from a known strain state, i.e., the stress-strain relation is implicit and costly to evaluate. ML-based methods promise a significant speed-up; given that a number of lower-scale simulations have been executed, i.e., a set of stress-strain data pairs is known, ML models can be used to interpolate the stress-strain data to arrive at an explicit model for the macroscopic stress-strain relation
\begin{equation}\label{eq:map-strain-stress}
\mathcal{M}: \bm{\varepsilon} \mapsto \bm{\sigma},
\end{equation}
which can then be used efficiently in forward simulations. Following the same idea, the data can alternatively be used to train a differentiable ML model to arrive at an explicit form of the strain energy density $W$
\begin{equation}\label{eq:map-strain-energy}
 \mathcal{M}: \bm{\varepsilon} \mapsto W.
\end{equation}
In this case, the stress-strain relation can be found by differentiating the above mapping with respect to the strain. A clear advantage of \eqref{eq:map-strain-energy} over \eqref{eq:map-strain-stress} is that the output of the mapping is scalar, which simplifies the training process. Expressions in the form of \eqref{eq:map-strain-energy} further simplify the enforcement of physical constraints such as thermodynamic consistency, as discussed in later sections.

\textbf{Uninterpretable ML approaches.} 
One of the first methods for the development of models in the form of \eqref{eq:map-strain-energy}, termed Numerically Explicit Potentials \cite{yvonnet_numerically_2009}, gathers data of RVE responses in a discretized loading space and interpolates the effective response with cubic splines \citep{yvonnet_numerically_2009} and later NNs \citep{le2015computational}. To reduce the computational cost associated with the micro-scale simulations for data generation, \cite{fritzen_--fly_2019} uses reduced-order RVE simulations to train a path-independent NN in the form of \eqref{eq:map-strain-stress}. The NN is then used as a surrogate for the macroscopic material behavior in forward finite element simulations, however, as the accuracy of the NN cannot be guaranteed outside the training domain, the authors propose to use reduced-order micro-scale simulations to predict the effective material response whenever predictions outside the training domain are required.

\cite{liu2019deep} proposes an ML-based multiscale material modeling method called Deep Material Network (DMN), which is constructed as a hierarchical network of layered composite building blocks with known analytical homogenization that maps the microscopic material properties $\mathbb{C}_{\text{micro}}$ to the effective material properties $\mathbb{C}_{\text{effective}}$
\begin{equation}
\mathcal{M}: \mathbb{C}_{\text{micro}} \mapsto \mathbb{C}_{\text{effective}}.
\end{equation}
Each building block takes as input the compliance matrices of the individual phases and outputs the effective material properties in a rotated coordinate system. By chaining the building blocks in a binary tree structure in such a way that consequent blocks take as input the compliance matrices of preceding blocks, the emerging network is able to encode the homogenization process of materials with complex heterogeneities. The network parameters are trained based on two- or three-dimensional \cite{liu2019exploring} linear elastic RVE simulations. Afterwards, in order to reduce the network complexity while retaining its predictive abilities, the authors suggest different model compression methods based on network node deletion and subtree merging. In \cite{liu_transfer_2019}, the same authors show that a pre-trained DMN can be used as initial input for the training process for new similar materials, e.g., the same type of composite but with different volume fraction $f$ of the material phases. In this way, the training process can be accelerated and it can be ensured that different networks share the same architecture. This in turn enables interpolation between trained networks: if e.g. two networks for volume fractions $f_1$ and $f_2$ are trained, the material behavior of a structure with volume fraction $f$, $f_1<f<f_2$, can be obtained from interpolation.
In \cite{gajek2020micromechanics} the properties of DMN are studied on a theoretical basis. To this end, the authors interpret the network as a hierarchy of homogenization functions of generalized standard materials. In this way, they show that the networks are thermodynamically consistent and exhibit stress-strain monotonicity. They further propose new types of building blocks by allowing the composite building blocks proposed by \cite{liu2019deep} to rotate, and discuss their implementation in forward finite element simulations in \cite{gajek_fedmn_2021}.
Further, \cite{nguyen_interaction-based_2021} and \cite{nguyen_micromechanics-based_2022} show applications of the DMN in the context of porous and woven materials, respectively.

Realizing that supervised training of NNs requires a high amount of stress-strain data pairs, which are hardly available from experimental measurements or too costly to generate from RVE simulations, \cite{man2011neural} proposes to use the energy-based characterization method (see \cite{furukawa_elastic_2008}) to train NN-based CLs in an unsupervised manner, i.e., by leveraging full-field displacement measurements and global reaction forces, but no stress data. \cite{huang_learning_2020}, \cite{liu_learning_2020} and \cite{liu_neural_2020} encounter the same problem and train NNs using indirectly measured data for the constitutive modeling of a fiber-reinforced plate (\cite{huang_learning_2020}) and two-dimensional in-plane shear material behavior (\cite{liu_learning_2020}, \cite{liu_neural_2020}).

Another context in which ML finds its application is that of reduced-order modeling. By solving reduced-order models at the micro-scale, the computational cost of two-scale simulations can be drastically decreased while retaining high computational accuracy. \cite{liu2016self} proposes to use ML, and in particular $k$-means cluster analysis, to divide the RVE geometry in the offline stage into a finite set of clusters with similar strain concentration. Assuming constant local variables over these clusters greatly simplifies the solution process of the Lippmann-Schwinger equation during the online stage. The method is non-parametric, naturally thermodynamically consistent, and is known as self-consistent clustering analysis. A great advantage of the method is that the offline stage can be computed under the assumption of small-strain elasticity, and the results of the offline stage can be utilized afterwards to compute material behavior beyond elasticity in the online stage. An equivalent approach is proposed independently by \cite{wulfinghoff_model_2018}, which derives the method from the perspective of the Hashin-Shtrikman variational principle. \cite{schneider_mathematical_2019} discusses the analogy between the methods proposed by \cite{liu2016self} and \cite{wulfinghoff_model_2018} from a theoretical perspective and further points out the connection between the self-consistent clustering analysis and a related reduced-order modeling method, i.e., the transformation field analysis. Applications and extensions of the self-consistent clustering analysis are provided by \citep{bessa2017framework,liu2018microstructural,yu2019self,zhang2019fast,cavaliere_efficient_2020,jaworek_hashin-shtrikman_2020,castrogiovanni_tfa_2021,waimann_microstructure_2021}. \cite{vlassis2023denoising} employ a denoising diffusion probabilistic model (DDPM) to map the reduced-order representation of small-strain elasticity to an embedded representation of the corresponding complex microstructures. Thus, they resolve the optimization  in a reduced-order space to facilitate the inverse design and generation of microstructures with specific targeted non-linear properties. 


Finally, we mention ML-based approaches which allow to construct black-box models $\mathcal{M}$ taking a set of microstructure or process parameters as input $\mathcal{I}$ and outputting a set of targeted effective material properties $\mathcal{O}$, e.g., the elastic stiffness
\begin{equation}
\mathcal{M}: \mathcal{I} \mapsto \mathcal{O},
\end{equation}
The model $\mathcal{M}$ may be any ML-based mapping, such as e.g. an NN or other function approximator. 
In this way, material properties may be predicted without experimental or numerical testing, but solely by knowing characteristic features of the production process and/or the microstructure. Interested readers are referred to the (non-comprehensive) list of references
\cite{tutumluer1998neural,
shin_identification_2003,
man2011neural,
asteris_feed-forward_2017,
yan_data-driven_2018,
huber_connections_2018,
yang2018deep,
messner_convolutional_2020,
rao2020three,
chen_learning_2021,
ni_deep_2021,
mianroodi_lossless_2021}.
The same idea can be extended to predicting mechanical properties as spatially varying fields. The ML-based mapping $\mathcal{M}$ may take spatial measurements $\mathcal{I}(\bm{x})$ as input  to predict a spatial property field $\mathcal{O}(\bm{x})$
\begin{equation}
\mathcal{M}: \mathcal{I}(\bm{x}) \mapsto \mathcal{O}(\bm{x}),
\end{equation}
where $\bm{x}$ denotes the spatial coordinate,
see e.g.
\cite{chen_learning_2021},
\cite{ni_deep_2021}.



\textbf{Model-free approaches.} A model-free approach is proposed in \cite{kirchdoerfer2016data} with a first application on truss systems. The data set is composed of stress and strain tensor pairs and the solution is obtained by minimizing a distance defined in an energetic sense between simulated and observed stresses and strains subject to kinematic and equilibrium constraints, the latter enforced through Lagrange multipliers. \cite{ibanez2018manifold} follows this idea by interpreting the given stress and strain data as a low-dimensional manifold that is embedded in the high-dimensional phase space (dimensionality is twelve, due to six independent components of the stress and strain tensors). \cite{leygue_data-based_2018,stainier2019model} propose the inverse problem of the problem posed by \cite{kirchdoerfer2016data,ibanez2018manifold}, see also \cite{stainier2019model,dalemat2019measuring}. The objective of their proposed method, denoted as DD identification, is to calculate the stress field given full-field displacement measurements, e.g., generated through DIC, without postulating any material model.



\subsection{Finite-strain elasticity}

Finite-strain elasticity, while sharing similarities with its small-strain counterpart, especially in the area of identifying elasticity parameters, introduces unique complexities.
A primary challenge resides in incorporating geometric nonlinearity and material frame considerations, which complicate the task of parameter identification \citep{holzapfel2002nonlinear}. 
Despite these complexities, constitutive modeling work has turned to DD models to facilitate and automate calibration of novel materials. 
These models do not merely aim for an accurate fit to the training data, but also for trustworthy predictions on unseen data points. 
The robustness of these predictions can be validated via thorough sensitivity checks of the input parameters, solidifying the reliability of the models for broader applications. 
In this section, we discuss the evolution of these models and the key considerations they embody, with a particular focus on parametrization and adherence to physical constraints.

Parametrization and the purposeful design of the input and output spaces are important aspects of DD finite-strain problems. 
The concern lies in the optimal determination of the input and output strain and stress states for the DD material laws. 
Some of the strain measures that are used in the finite-strain description are the deformation gradient ($\boldsymbol{F}$), the Green-Lagrange strain tensor ($\boldsymbol{E}$), and the right Cauchy–Green deformation tensor ($\boldsymbol{C}$), paired with their respective conjugate stress measures: the first Piola-Kirchhoff stress ($\boldsymbol{P}$) or the second Piola-Kirchhoff stress ($\boldsymbol{S}$).
In the case of hyperelasticity, the formulation involves learning energy density functions with respect to these variables, as illustrated by
\begin{equation}
\boldsymbol{P}(\boldsymbol{F}) = \frac{\partial \bar{W}(\boldsymbol{F}) }{\partial \boldsymbol{F}} \quad \text{or} \quad
\boldsymbol{S}(\boldsymbol{E}) = \frac{\partial \hat{W}(\boldsymbol{E}) }{\partial \boldsymbol{E}} = 2 \frac{\partial \tilde{W}(\boldsymbol{C}) }{\partial \boldsymbol{C}}.
\end{equation}
The frame of reference chosen for parametrization has a substantial impact on the size and complexity of the learning problem. For instance, the deformation gradient $\boldsymbol{F}$ and the stress tensor $\boldsymbol{P}$ consist of nine components as two-point tensors, while the tensors $\boldsymbol{E}$, $\boldsymbol{C}$, and $\boldsymbol{S}$ are symmetric with six components each. 
This difference calls for close consideration as it involves learning a non-symmetric fourth-order elasticity tensor $\mathbb{C}^{\boldsymbol{PF}}(\boldsymbol{F})$ (with 81 coefficients in general) as opposed to symmetric $\mathbb{C}^{\boldsymbol{SE}}(\boldsymbol{E}), \mathbb{C}^{\boldsymbol{SC}}(\boldsymbol{C})$ elasticity tensors, both of which possess major and minor symmetries and thus require calibration for fewer elasticity coefficients. 
The CL can be formulated in terms of strain energy density as 
\begin{equation}\label{eq:map-strain-energy-finite}
\mathcal{M}: \boldsymbol{F} \mapsto \bar{W}, \quad
\mathcal{M}: \boldsymbol{E} \mapsto \hat{W} \quad \text{or} \quad \mathcal{M}: \boldsymbol{C} \mapsto \tilde{W},
\end{equation}
or relying on the corresponding strain-stress mappings, i.e.
\begin{equation}\label{eq:map-strain-stress-finite}
\mathcal{M}: \boldsymbol{F} \mapsto \boldsymbol{P}\,, \quad  \mathcal{M}: \boldsymbol{C} \mapsto \boldsymbol{S}\,, \quad \text{or} \quad \mathcal{M}: \boldsymbol{E} \mapsto \boldsymbol{S}.
\end{equation}
Formulations also often involve mappings from invariants of strain tensors to energy and stress measures, as addressed in numerous research works discussed below. 
Either of the aforementioned finite-strain elastic tensors depend on the current state of deformation, making their definition more complex than in the small-strain framework where only one constant tensor is needed. An additional important property of hyperelastic material models is polyconvexity, which along with coercivity is a sufficient condition for the solvability of boundary value problems under general boundary conditions and body forces. 

The selection of strain and stress inputs and outputs in DD laws has consequences on an important required property of finite-strain material laws, namely, material frame indifference/objectivity. This property ensures that both the energy and stress response of the material remain constant upon rigid body motion and rotation. 
In the context of hyperelasticity, it is defined as follows
\begin{equation}
\bar{W}(\boldsymbol{F}) = \bar{W}(\boldsymbol{QF}), \quad
\bar{\boldsymbol{P}}(\boldsymbol{F}) = \bar{\boldsymbol{P}}(\boldsymbol{QF}), \quad
\forall \boldsymbol{Q} \in SO(3),
\label{eq:frame_invariance_F}
\end{equation}
where $\boldsymbol{Q}$ represents a rotation tensor and $SO(3)$ denotes the 3D rotation group. DD material laws expressed in terms of $\boldsymbol{P}-\boldsymbol{F}$ relations do not necessarily satisfy objectivity and may require additional enforcement of the related constraint, whereas material laws expressed as $\boldsymbol{S}-\boldsymbol{E}$  or $\boldsymbol{S}-\boldsymbol{C}$ relations are known to be automatically objective \cite{holzapfel2002nonlinear}.
Frame invariance can also be inherently achieved by selecting an appropriate strain invariant formulation. 

\textbf{Uninterpretable ML approaches.} One of the earliest attempts to model finite-strain elasticity using NNs goes back to  \cite{shen2004neural,liang2008neural}. 
These authors mainly focus on rubber materials, utilizing a small NN to model a hyperelastic energy function. 
Their approach resembles the finite-strain model proposed by \cite{ogden1997non}, which also serves as a benchmark.
The authors adopt a strain invariant formulation of the input, ensuring the concise representation of the strain characteristics. 
Moreover, their approach also emphasizes the need for stress and higher-order derivatives from the learned function, particularly to facilitate the successful implementation of the trained NN within the finite element solver. 
This early exploration of finite-strain elasticity via NNs showcases the potential and challenges inherent to the task.

More recent work in finite-strain elasticity further explores the question of training a hyperelastic energy function and subsequently deriving the stress information through differentiation. 
This preference is rooted not only in its potential to simplify the learning problem – mapping from the strain space to a single, scalar-valued function $W$ – but also in its utility in enforcing and validating necessary properties in the learned material law.
\cite{sagiyama2019machine} leverages both feed forward NNs and CNNs for predicting a homogenized energy function of a single microstructure for data from multiple numerical simulations of evolving crystal microstructures. The trained NNs then facilitate the calculation of the  $\boldsymbol{P}(\boldsymbol{F})$ values through differentiation of the learned $W(\boldsymbol{F})$ in the multiscale approach.
Meanwhile, \cite{vlassis2020geometric} introduces the Sobolev training technique by \cite{czarnecki_sobolev_2017} in the hyperelasticity training procedure for polycrystals.
By doing so, the authors are able to constrain the optimized architecture to accurately predict derivatives of the learned $W(\boldsymbol{C})$ to calculate $\boldsymbol{S}$. Furthermore, by adopting a graph representation of the polycrystals as material descriptors of anisotropy, they also generalize the law to a family of microstructures.
\cite{klein2022polyconvex} encodes hyperelastic energy functions using input convex NNs, addressing the issues of polyconvexity and thus material stability. They explore two methods for obtaining polyconvexity: firstly, they employ polyconvex, anisotropic, and objective invariants as inputs; secondly, they use the deformation gradient, its cofactor, and determinant as input while performing additional data augmentation to satisfy the objectivity condition.
\cite{as2022mechanics} also discusses an input convex NN-based energy function regression model framework to enforce mechanics principles such as objectivity, consistency, and dynamic material stability through proper parameterization of the energy mapping in the $\boldsymbol{S}-\boldsymbol{E}$ frame, augmenting the loss function, and modeling the NN weights after a softplus function respectively. 
Along similar lines, \cite{thakolkaran2022nn} trains input convex NNs as hyperelastic strain energy density functions and satisfies key physical constraints through a specifically designed NN architecture. In this work, training is carried out in an unsupervised manner, i.e. using a loss function which enforces balance of linear momentum based only on full-field displacement and global force data, and no stress data. Hence, the approach is directly applicable to experimental measurements (see also the discussion on interpretable approaches based on sparse regression).
\cite{tac2022data} employs neural ordinary differential equations (ODEs), a form of polyconvex NNs, to develop DD material models that inherently satisfy the polyconvexity condition in elasticity, using the properties of ODEs to create monotonic functions that approximate the strain energy derivatives, thus 
effectively modeling highly nonlinear, anisotropic materials.
\cite{vlassis2021md} trains anisotropic NN hyperelastic models for monoclinic crystals in terms of both $\boldsymbol{P}-\boldsymbol{F}$ and $\boldsymbol{S}-\boldsymbol{E}$ relations. This study also incorporates energy and stress frame invariance terms into the loss function during training and discusses post hoc validation tests to test for material stability.
\cite{fernandez2021anisotropic} deploys NNs to learn the homogenized three-dimensional constitutive behavior of anisotropic hyperelastic cubic lattice metamaterials while ensuring objectivity and material symmetry in the problem formulation. 
Expanding on this approach, \cite{fernandez2022material} embeds a parametric dependence in the NN formulation, which allows to capture intricate topological and material variations, thus refining the learned energy functions.

Many ML studies in the literature establish a direct mapping from the strain to the stress material response. This strain-stress approach is sometimes preferred over an energy formulation due to its more straightforward implementation and the common lack of energy data samples. 
\cite{yang2019derivation} utilizes principal stress-stretch data in the training of elastic NN laws, which are found to approximately satisfy objectivity conditions.
In contrast, \cite{chung2021neural} utilizes NNs to formulate CLs in terms of $\boldsymbol{S}-\boldsymbol{E}$ using a dataset derived from molecular dynamics, later validating the learned constitutive tangent modulus through numerical differentiation.
In another study, \cite{im2021neural} NNs are trained to establish both the stress-strain $\boldsymbol{S}-\boldsymbol{E}$ and stiffness-strain $\mathbb{C}^{\boldsymbol{SE}}-\boldsymbol{E}$ nonlinear relations for crystal structures under various symmetry conditions. \cite{fuhg2021local} conducts a comparative analysis of NNs, with and without Sobolev constraints, and local approximate Gaussian process regression in mapping stress-strain $\boldsymbol{S}-\boldsymbol{C}$ and employ Latin hypercube sampling, c.f. section \ref{sec::DataSampling},  to uniformly sample the space spanned by the deformation gradient.

\textbf{Interpretable ML approaches.} 
\cite{latorre2020experimental} models hyperelastic behavior for soft materials by identifying the energy function using spline regression and smoothing penalization, therefore producing an interpretable ML model. Notably, the authors use stability conditions in order to handle the noisy data often present in such materials. Also, \cite{frankel2020tensor} and \cite{fuhg2022physics} use the representation theorem of tensor-valued tensor functions, c.f. \cite{haupt2002continuum}, which is used to write the stress as a linear combination of invariant-dependent coefficient functions and basis generators. This allows learning the mapping from the invariants of the deformation to the values of the coefficients and guarantees that the material response is frame indifferent. In this context, \cite{fuhg2023stress} studies and compares different formulations of the representation. In order to lower the number of required training samples \cite{fuhg2022physics} proposes a space-filling sampling algorithm in invariant space.

Recently, several efforts have been made to deduce analytical expressions for material models from data using either symbolic regression or sparse regression (see the discussion on terminology in Section \ref{ML_approaches}).
In the context of hyperelasticity, symbolic regression has been used since the early work by \cite{schoenauer_evolutionary_1996}, see \cite{abdusalamov_automatic_2023} for a more recent application. Even more recently, \cite{Kissas2024} proposes a symbolic regression method in which potential expressions for physically valid hyperelastic CLs are generated using regular tree grammars and model discovery is carried out by combining variational autoencoders and a covariance matrix adaptation evolutionary strategy.

Sparse regression in the context of material modeling on the other hand is a very recent research field.
The first works in which sparse regression from a potentially large library of candidate material models is used to discover hyperelastic strain energy functions as symbolic mathematical expressions are \cite{flaschel_unsupervised_2021}, that develops a method denoted as EUCLID standing for Efficient Unsupervised Constitutive Law Identification and Discovery (see \cite{flaschel_automated_2023} for an overview), and \cite{wang_inference_2021}, that develops a method coined Variational System Identification.
Beside being interpretable approaches, these methods are also unsupervised.
Instead of relying on labeled stress-strain data pairs, which are in reality only available under very simple loading conditions like uniaxial tension or simple torsion, the material model discovery is informed by experimentally measurable displacement and global reaction force data.
The lack of stress labels is compensated for by applying a physics-motivated loss function based on the conservation of linear momentum.
This has the advantage that the learning process can be informed by real data instead of data generated from micro-scale simulations, which are only feasible if the microstructure of the material and models of all its material constituents are known.
It is shown by \cite{joshi2022bayesian} that the EUCLID framework can also be considered from a Bayesian perspective for discovering symbolic hyperelastic models with quantifiable uncertainties.
A first experimental validation of a supervised version of EUCLID is provided in \cite{flaschel_automated_2023-2}, where sparse regression is leveraged to discover hyperelastic strain energy functions for human brain tissue using data stemming from uniaxial tension and torsion tests of human brain tissue.
In \cite{Kissas2024}, the hand-crafting of the material model library required in sparse regression approaches is replaced by its automatic generation as the "language of hyperelastic material models", i.e. as the set of expressions generated by a context free grammar while accounting for physics constraints.
\cite{linka2023automated} discusses the discovery and interpolation of hyperelastic models for the human brain tissue in a framework that selects model inputs and forms from a family of constitutive building blocks of classical models from the hyperelasticity literature through the optimization of an NN. Another approach for replacing the hand-crafting of libraries is suggested in \cite{fuhg2024extreme} by enabling extreme sparsification of physics-augmented neural networks utilizing a smoothed $L_0$ regularization approach showcasing a combination of expressivity and interpretability in a range of experimental and synthetic test cases.

\textbf{Model-free approaches.} The strain-stress formulation is particularly appealing for model-free applications.
For instance, in \cite{nguyen2018data}, a model-free approach is extended from \cite{kirchdoerfer2016data} to finite-strain nonlinear elasticity.
The physical constraints (the principle of virtual work) are enforced with Lagrange multipliers and the search problem is formulated in terms of $\boldsymbol{S}-\boldsymbol{E}$. This choice is made to automatically satisfy the objectivity conditions and ensure the symmetry of the stiffness matrix.
Also, \cite{conti2020data} studies well-posedness and existence of minimizers of model-free finite strain formulations and proposes a formulation of finite elasticity in terms of $\boldsymbol{P}-\boldsymbol{F}$. It provides the necessary equilibrium and compatibility constraints, defines conditions for material frame indifference in this approach, and delves into the concepts of convexity within the finite-strain model-free framework.
Moreover, \cite{he2020physics} employs a DD approach to model anisotropic nonlinear elasticity. The solution search is conducted in both strain-stress data and anisotropy orientation spaces based on manifold embedding and a convexity-preserving reconstruction scheme, called local convexity DD computing.

%
The model-free approach based on distance minimization \cite{kirchdoerfer2016data} is extended to account for finite deformations by introducing appropriate metrics in $(\boldsymbol{F}, \boldsymbol{P})$ or $(\boldsymbol{E}, \boldsymbol{S})$ phase spaces, as discussed by \cite{nguyen2018data} and \cite{platzer2021finite}. To enhance the performance of the model-free solver in scenarios with limited or noisy data, \cite{he2020physics} and \citep{he2021manifold} incorporate an online locally linear embedding scheme, leveraging the advantages of local convex linear interpolation. Additionally, \cite{he2021deep} develop an autoencoder framework to enhance local search efficiency and mitigate noise sensitivity within the model-free approach by discovering lower-dimensional embedding spaces. \cite{bahmani2022manifold} introduces a global manifold learning approach employing invertible NNs, allowing for direct interpolation on the manifold and eliminating the need for local discrete searches. This concept is extended to incorporate an isometric (distance-preservation) constraint to maintain metric structures between ambient and embedding phase spaces \cite{bahmani2023distance}. This method introduces a geometrically inspired regularization technique within the classical autoencoder framework, facilitating noise reduction and interpolation on the data manifold.


\section{DD modeling for path-dependent CLs}\label{sec::intro}

This section is devoted to review the DD approaches to describe path-dependent materials. The characteristic feature of this class of materials is that the current stress depends on both the current strain and on the entire deformation history \cite{Rivlin1972}. Therefore, the bijectivity of the CL is not fulfilled since a single state of deformation can pair with a possibly infinite number of stress states, leading to the so-called \textit{one-to-many} strain-stress mapping. This is in contrast with the path-independent models, where  each strain state corresponds to only one stress state, leading to a \textit{one-to-one} mapping. The description of the path-dependent constitutive behavior is also often complicated by an irreversibility constraint of the (dissipative) processes characterizing the material response. One potential way of addressing the one-to-many mapping issue is to introduce in the CL so-called history variables, also known as internal variables (since they are not directly observable), whose definition should ensure a one-to-one mapping, i.e. each strain-stress pair corresponds to a unique set of internal variables.

In the following we specifically focus our attention on hypo- and elastoplasticity, viscoelasticity, damage, fracture, fatigue and, finally, on their coupling with other physical processes, i.e. within a multiphysics representation of the constitutive behavior. As in Section~\ref{sct:pathindep}, we begin each section with a brief summary of the main features of the specific material behavior and then we review the most significant contributions in the context of DD approches.


\subsection{Plasticity}
The perhaps most widely used history-dependent material modeling framework is plasticity. 
Over the years, different theories of plasticity have been developed, such as hypoplasticity \cite{wu2000hypoplasticity}, elastoplasticity \cite{hill1998mathematical}, hyperplasticity \cite{houlsby2007principles} or generalized plasticity \cite{yu2006generalized}. Small-strain elastoplasticity makes a distinction between elastic and inelastic components of the constitutive response by an additive decomposition of the strain tensor into elastic and plastic strains and introduces a notion of yield surface $f(\bm{\sigma}, \bullet)$, which is dependent on a stress measure $\bm{\sigma}$ and on a number of thermodynamic forces used to define the hardening behavior. The yield function defines the plastically admissible domain as the set of stresses for which $f(\bm{\sigma}, \bullet) \leq 0$. 
The yield function is classically decomposed as $f = \overline{\sigma}(\bm{\sigma}, \bullet) - k(\bullet) $ into a scalar \textit{equivalent stress} $\overline{\sigma}$, that can be e.g. dependent on a kinematic hardening tensor, and a 
resistance $k$, often 
a function of a \textit{drag stress} which is used to describe isotropic hardening.
On the other hand hypoplastic models, which are commonly applied in geomechanics, often do not separate the strain into reversible and irreversible parts and do not employ yield functions to characterize the onset of yielding. Here, the information about the past is all concentrated in the current stress.
Mirroring this distinction, as follows we classify DD constitutive models for plasticity into models utilizing hypoplastic and elastoplastic ideas.

\subsubsection{DD plasticity modeled after hypoplasticity }\label{hypoplasticity}
We start with the hypoplastic case which, for isotropic materials, aims to predict the current stress or strain values from a time-discrete version of the CL
\begin{equation}
\mathcal{M}: \{\bm{\sigma}, \dot{\bm{\varepsilon}}, \bullet\} \mapsto \dot{\bm{\sigma}}     
\end{equation}
where $\dot{\bm{\sigma}}$ is the Jaumann stress rate and $\mathcal{M}$ is a tensorial function \citep{wu1994simple}. 

\begin{figure}[h!]
\begin{subfigure}[b]{0.5\linewidth}
        \centering
    \includegraphics[scale=0.3]{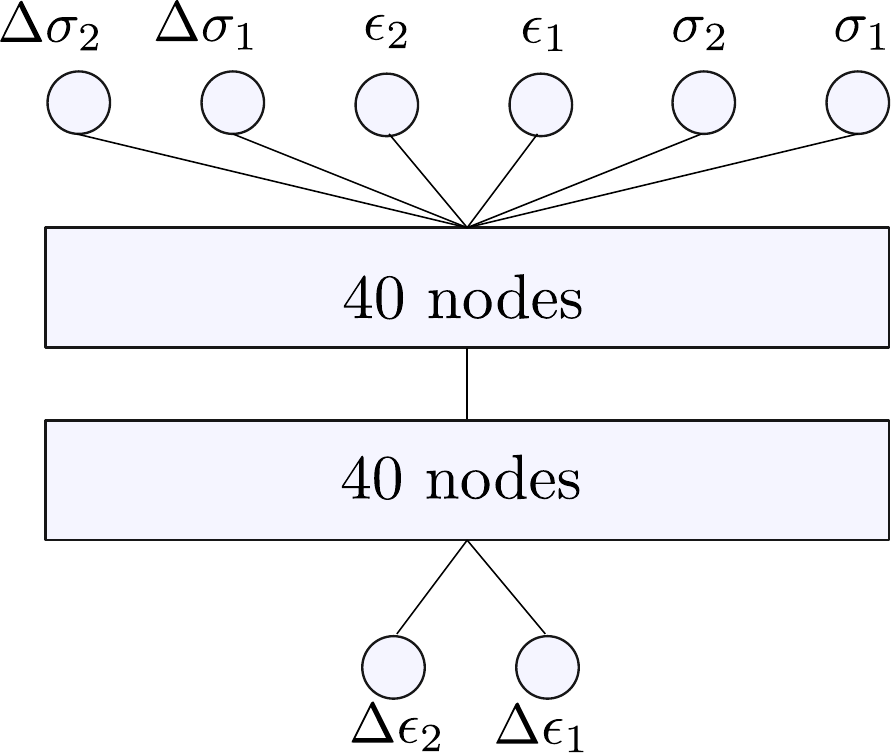}
    \subcaption{Biaxial monotonic loading}\label{fig::ghabFirstMono}
\end{subfigure}
\begin{subfigure}[b]{0.5\linewidth}
        \centering
    \includegraphics[scale=0.3]{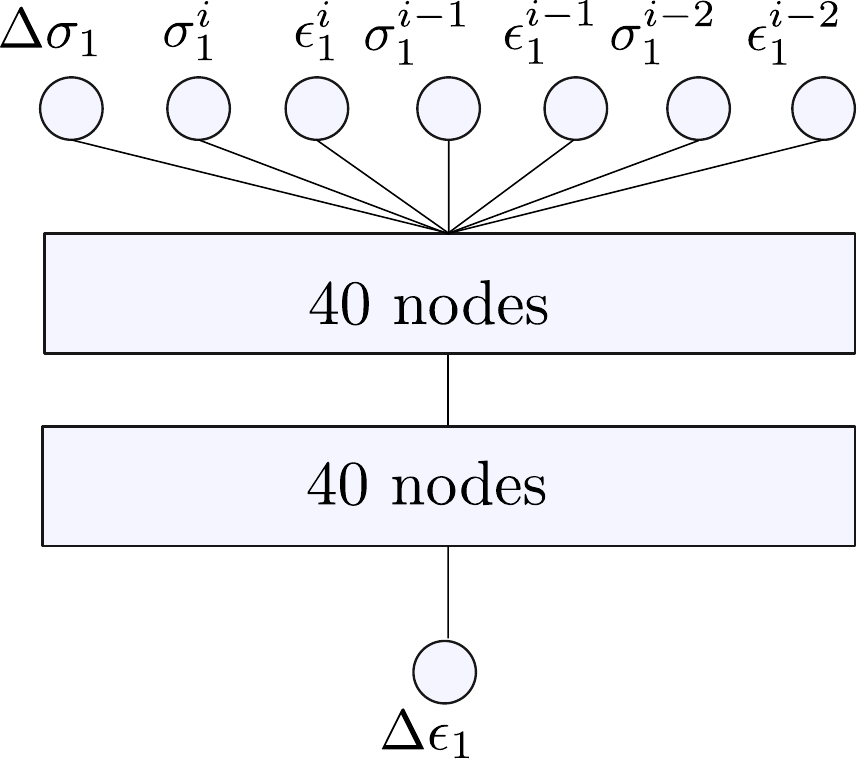}
    \caption{Uniaxial cyclic loading}\label{fig::ghabFirstCyclic}
\end{subfigure}
    \caption{First NN architectures based on quasi-sequential data used for stress-controlled path-dependent modeling of concrete inspired by \cite{ghaboussi1991knowledge}. }
\end{figure}

\textbf{Uninterpretable ML approaches.} Since the early works on ML plasticity modeling were performed in the geomechanics community, they have close similarities to the hypoplastic theory. 
These include the first and perhaps most influential ML constitutive models for history-dependent problems published in the early 1990s by the group of J. Ghaboussi.
The first study employs a standard NN with sigmoidal activation functions to fit monotonic loading data from plain concrete under a biaxial state of stress at small strains \citep{ghaboussi1990material}.
By treating the mapping problem in a quasi-sequential manner, this initial study already introduces most of the major ideas that will be found in papers of the following years. Both stress- and strain-controlled models are developed.
Both approaches in this setting require a network with six inputs and two outputs; e.g. in the stress-controlled approach, the six inputs are the two principal stresses, the two principal strains and two stress increments ($\Delta \sigma_{1}$, $\Delta \sigma_{2}$), whereas two strain increments ($\Delta \epsilon_{1}$, $\Delta \epsilon_{2}$) are given as outputs.  The utilized NN architecture is visualized in Figure \ref{fig::ghabFirstMono}.
The authors only offer visual comparisons between experimental and approximated responses on selected test loading paths, and
the developed DD model (strain-controlled) is not tested in a finite element framework.
The approach is extended to uniaxial cyclic loading in a follow-up paper
\citep{wu1990representation}. Here the architecture of the network is (heuristically) changed by the authors to be able to distinguish loading and unloading curves by not only using the current point on the stress-strain curve as input but also utilizing the previous two points on the loading path, see Figure \ref{fig::ghabFirstCyclic}.
More information on the networks, as well as the modeling ideas, can be found in \citep{wu1991neural}. The findings are reiterated in \cite{ghaboussi1992potential,ghaboussi1992neuro}.

A similar approach as in \citep{ghaboussi1990material} (training history-dependence in a hypoplastic framework) for DD material modeling of sand is tested in \cite{ellis1992neural}, where an additional material parameter is considered as input.
In the 1990s, various authors reapply the same ideas to laminates \cite{pidaparti1993material}, clays \citep{penumadu1994rate,amorosi1996use}, sand \citep{ghaboussi1994neural,sikora1998neural}, rocks \citep{millar1994investigation,millar1995investigation}, soft soils \citep{logar1997neural} or concrete \citep{wu1993modelling}.
This approach is also studied and tested in a more rigorous way on different material behaviors in \cite{pernot1999application}.
In contrast to these works, the study in
\cite{Hadjigeorgiou1995} defines two NNs of the same architecture to model rock joint material behavior; one is used to deal with small-displacement behavior and the other one with large displacements (the value of $1.5$ mm separating the two ranges is arbitrarily chosen by the authors). This approach aims to distribute the nonlinearities of the material model between the two networks (since the networks at this point cannot be too deep due to the available computational resources).
Other than NNs, SVR-based plastic modeling of geomaterials is investigated in \cite{zhao2014simulating,shen2015least} while polynomial regression models are studied in \cite{javadi2009applications,javadi2009intelligent,faramarzi2014epr}.
All of these follow the same idea of using time-discrete current strain and stress values as inputs and the predicted stress as an output to their formulations.
An adaptive approach to determine the ideal network architecture of the NNs for path-dependent material modeling is proposed in \cite{wu1991neural,joghataie1995learning,ghaboussi1997nested,ghaboussi1998new}. The number of nodes and the necessary number of history inputs, i.e. $([\epsilon_{j}, \sigma_{j}], [\epsilon_{j-1}, \sigma_{j-1}], \ldots, )$, can be trained adaptively. The authors use the term "nested" NN for this process, which is later used for modeling the behavior of soils \cite{fu2007integration}.
This technique is also employed to obtain path-dependent ML CLs from indirect data, such as global load-deflection responses or displacement fields. Here, the training algorithm of the ML tool is built around a numerical framework such as the finite element method (FEM)\cite{ghaboussi1998autoprogressive}. Other works employ similar approaches in later years
\cite{sidarta1998constitutive,shin2000self,pande2002finite, hashash2002systematic, jung2006characterizing, yun2012improved}.
However, approaches that adaptively change the architecture of NNs seem to have been no longer actively pursued in the context of material modeling.

The approaches reviewed so far are based on using the history of observable quantities such as stress and strain as inputs to a ML model in order to predict the next stress state. 
However, due to the path-dependence of the stress evolution, these methods likely fail when studying longer loading curves with complex loading patterns or when extrapolating outside of the domain of the training data. To address this issue, more advanced ML approaches introduce internal variables as additional inputs. In particular, \cite{furukawa1997neural,furukawa1998implicit} include for the first time internal variables in a ML stress evolution law. 
They focus on a viscoplastic model with kinematic and isotropic hardening and utilize an NN, with the current strain, the internal variables $\bm{\xi}$ and the current stress as input, and the current rates of the viscoplastic strain $\bm{\varepsilon}_{\text{vp}}$ and of the internal variables as outputs. For this purpose, the authors define an implicit constitutive model in the state space of the form
\begin{equation}
\mathcal{M}: \{\bm{\varepsilon}_{\text{vp}}, \bm{\xi}, \bm{\sigma}\} \mapsto \{\dot{\bm{\varepsilon}}_{\text{vp}},\,  \dot{\bm{\xi}}\} 
\end{equation}
Assuming that the initial conditions 
are known, a forward Euler scheme is employed to update the variables from time step $n$ to $n+1$ with
\begin{equation}
\begin{aligned}
\bm{\varepsilon}_{\text{vp},n+1} &= \bm{\varepsilon}_{\text{vp},n} + \dot{\bm{\varepsilon}}_{\text{vp},n} \Delta t \\
\bm{\xi}_{n+1} &= \bm{\xi}_{n} + \dot{\bm{\xi}} \Delta t .
\end{aligned}
\end{equation}
The stress update is obtained in a similar manner. In this approach, the internal variables have to be explicitly known. If this is not the case, the authors describe a procedure to obtain information about the internal variables from cyclic loading curves and apply this methodology in an example.

Other work dealing with the explicit definition of the internal variables includes \cite{haj1998micromechanics, haj2001simulated}
which rely on data from repeating unit cells.
In \cite{yun2008new}, an energy-based internal variable and an internal variable that implies the direction for the next time-
or load-step along the equilibrium path are used as additional inputs to an NN.
Similar approaches that make use of known or implicitly obtained internal variables as inputs to NNs that describe the stress evolution law can be found later in \cite{huang2020machine}, where the accumulated absolute strain is used as an additional input.
Nowadays, the use of internal variables as additional inputs to guide the training process is becoming more and more established as DD modeling of path-dependent materials shifts more and more away from experimental data towards data obtained from numerical simulations.
One of the problems associated with using plain NN approaches that adopt previous time-discrete strains and stresses as input and current stress as output is their dependence on the strain increment.
In order to avoid this problem, \cite{Unger2008} proposes a two-stage algorithm based on the total strains for modeling the homogenized response of a lower-scale problem.  In the first step, a SVM is used to classify whether the current loading step (described by total strain and last equilibrium value of the history variable) results in an update of the history variable, which is comparable to a return mapping procedure. Based on this information, in the second step, the stresses are obtained from a NN.   

\begin{figure}[h!]
    \centering
    \includegraphics[scale=0.9]{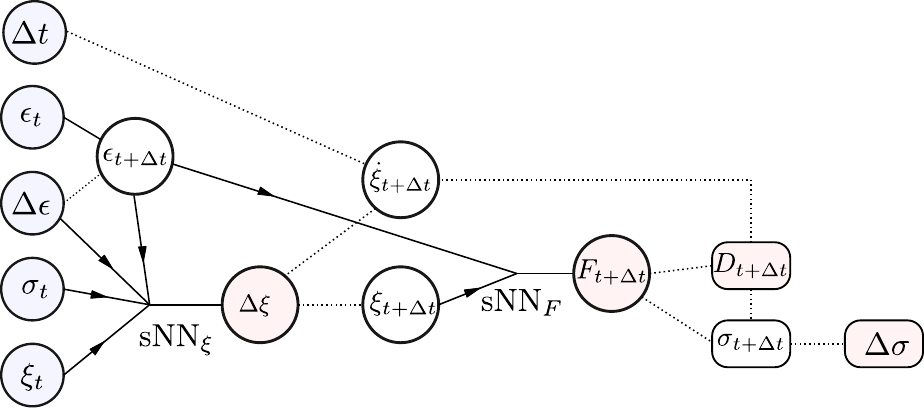}
    \caption{TANNs proposed by \cite{masi2021thermodynamics} 
    }
    \label{fig:TANNMasi}
\end{figure}

The aforementioned approaches suffer of reliability and robustness issues when the behavior need to be extrapolated outside of the range of data used during the training. A possibility to relieve this drawback is to introduce physical consistency into the training process, a technique that can also reduce the data hunger of the method.
To the best of the authors' knowledge, \cite{lefik2003artificial} were the first to include a physical concept in the training process of an NN used for DD path-dependent material modeling; they enforce isotropy of their CL in a soft manner by augmenting the training dataset by rotated counterparts of existing data. The concept of physics-informed ML in the incremental modeling of path-dependent materials using NNs is recently receiving increasing attention, with two recent works leading the way. First of all, \cite{xu2021learning} propose using the Cholesky factor $L$ of the tangent stiffness matrix as the NN output instead of the stress.
The stress update is obtained by
\begin{equation}
    \bm{\sigma}_{n+1} = L(\bm{\varepsilon}_{n+1}, \bm{\varepsilon}_{n}, \bm{\sigma}_{n}) L(\bm{\varepsilon}_{n+1}, \bm{\varepsilon}_{n}, \bm{\sigma}_{n})^{T} (\bm{\varepsilon}_{n+1} - \bm{\varepsilon}_{n})  + \bm{\sigma}_{n}.
\end{equation}
The positive aspect of this formulation is that it implicitly enforces the tangent stiffness matrix to be symmetric semidefinite.
Secondly, thermodynamics-based NNs (TANNs) encode the underlying thermodynamic principles directly into the networks \cite{masi2021thermodynamics,masi2022multiscale}. This is achieved by relying on a dual potential formulation, consisting of the Helmholtz free energy density $F$ and the dissipation rate potential $D$, from which the variables of interest can be derived in a thermodynamically consistent form.
In the isothermal case, the framework consists of two NNs, denoted as $\text{sNN}_{\xi}$ and $\text{sNN}_{F}$. The former is used to predict the increment of the internal variables
\begin{equation}
    \Delta \bm{\xi} = \text{sNN}_{\bm{\xi}} (\bm{\varepsilon}_{n+1}, \Delta \bm{\varepsilon}_{n}, \bm{\sigma}_{n}, \bm{\xi}_{n})
\end{equation}
The other network predicts the value of the Helmholtz free energy density of the next time step
\begin{equation}
   F_{n+1} = \text{sNN}_{F} (\bm{\varepsilon}_{n+1}, \bm{\xi}_{n+1})
\end{equation}
which can be used to obtain the stress update with $\bm{\sigma}_{n+1} = \frac{\partial F_{n+1}}{\partial \bm{\varepsilon}_{n+1}}$. The TANN architecture is schematized in Figure \ref{fig:TANNMasi}.
As a result of the (implicit) thermodynamic consistency, the task of identifying the underlying pattern of thermodynamic laws no longer needs to be performed by the ML tool. 

A possibility to deal with path-dependence without explicitly introducing internal variables is relying on ML architectures that embed and describe internal states. In this context, RNNs have been investigated in the literature for plastic material modeling. The study in \cite{ellis1995stress} is the first to apply sequential NNs, specifically the Jordan network \citep{jordan1990attractor}, in order to learn the path-dependent stress-strain curve of sand under consideration of a constant strain increment in a triaxial state of stress. 
In contrast to the previous approaches, the internal variables are here implicitly stored in the NN.
Based on investigations of \cite{ellis1995stress} this type of NN offers better results than simple NNs for the application at hand. For this reason, they utilize a network with seven inputs, including the current stress and strain rates, while the stress state resulting from the next strain increment defines two network outputs.
Some critical points of this paper are further discussed in \cite{najjar1996discussion}, including the issue that the network is only applicable to a specific strain rate.
The same network architecture is later used by
\cite{pendmadu1997} to model clay.
The same authors \citep{penumadu1999triaxial,penumadu2000virtual} later improve upon this approach to build a ML model for sand and gravel by (similarly to \cite{Hadjigeorgiou1995}) dividing the input space of a feature into two parts, in order to train two different sequential NNs for the same dataset. 
Instead of using Jordan networks, the studies in 
\cite{zhu1998modeling,zhu1998modelling} employ Elman networks \citep{elman1990finding}, i.e. another simple RNN, for simulating and
predicting the shear behavior of two different soils.
Simple RNN architectures are also used to model unsaturated soils \cite{habibagahi2003neural}, clays \cite{najjar2007simulating} or sand \cite{romo2001recurrent}. In
\cite{najjar2002simulating} a sequential NN is employed for the three-dimensional stress-strain relationship of sand under monotonic loading.
Different mapping techniques including quasi-sequential mapping, sequential mapping, function fragmentation, and function labeling for history-dependent  material datasets are compared in \cite{basheer1998modeling} and \cite{basheer2000selection}, where a hybrid quasi-sequential approach (relying on function labeling) is found to perform most proficiently. Later on, the author also tests time-delay NNs for history-dependent datasets \citep{basheer2002stress}.

Recently, other types of 
RNNs such as
LSTMs and GRUs have received more attention
\cite{oeser2009modeling,zopf2017numerical, Wang2018a, qu2021towards,ghavamian2019accelerating, gorji2020potential,wu2020recurrent, abueidda2021deep, chen2021deep, bonatti2022cp} because they are able to be trained more reliably by avoiding the vanishing/exploding gradients problems that are associated with the earlier version of RNNs. 
Even though these general DD frameworks for path-dependent materials are conceptually similar to the early approach proposed by \cite{ellis1995stress}, they are now predominantly used for three-dimensional applications and also include other parametrizations as input. 
E.g. the study in \cite{mozaffar2019deep} describes the sequential nature of plasticity using a GRU 
which takes the history of spatially averaged strains, material properties for each microstructural
phase and the current time as input and maps them to the predicted spatially averaged stress.
Other work includes
\cite{frankel2019predicting}, where an
LSTM takes the strain history over a sequence of times as a vectorial input as well as the latent space data from three-dimensional crystallographic orientation images using a CNN. The LSTM is then used to predict the stress evolution.
This idea is later refined by the same authors \cite{frankel2020prediction} by relying on
convolutional LSTMs to resolve the stress evolution also spatially.
In order to reduce the reliance on user-chosen hyperparameters, 
\cite{fuchs2021dnn2} proposes an approach based on RL that finds the optimal hyperparameter settings (including the network architecture) of RNNs for modeling path-dependent materials. 
Lastly, \cite{stocker2022novel} defines an
adversarial training scheme based on input perturbations that increases the prediction robustness of models trained with RNNs (GRU specifically).
A general problem associated with RNNs (such as LSTMs or GRU) in the training of path-dependent plastic material behavior is the dependence of these methods on the size of the increment. To overcome this challenge, \cite{bonatti2022importance} proposes a new RNN architecture that enforces self-consistency, i.e. their predictions converge when the increment size is decreased.
A different approach is proposed in \cite{liu2021learning}, which treats the material model prediction as a mapping between two function spaces without the need for time discretization.

As a remark, we could not find any ML path-dependent material modeling approach based on RNNs that includes physical concepts in the training procedure. This is definitely an area where more work and development is necessary.
The work that comes the closest is the one in 
\cite{jones2021neural}, where a neural ODE framework is employed to model the stress evolution. In contrast to RNNs, a neural ODE incorporates time step
scaling of the dynamics. Using this architecture as an internal state variable model and relying on representation theory, the authors are able to build a ML model that obeys physical principles, such as frame invariance and the second law of thermodynamics.

\textbf{Model-free approaches.} The framework based on the model-free paradigm \cite{kirchdoerfer2016data} is more complicated when dealing with path-dependent material behavior. In \cite{eggersmann2019model}, the material dataset is enlarged with internal variables. In \cite{karapiperis2021data} an energy-based parametrization is proposed that augments the phase space with the free energy and the dissipation, thereby enforcing thermodynamic consistency at all times. Similar approaches include \cite{ibanez2018manifold, ladeveze2019data}.
A different model-free idea is studied in
\cite{tang2020map123,tang2021map123}, where the three-dimensional stress/strain state is projected onto uniaxial tension/compression data. 

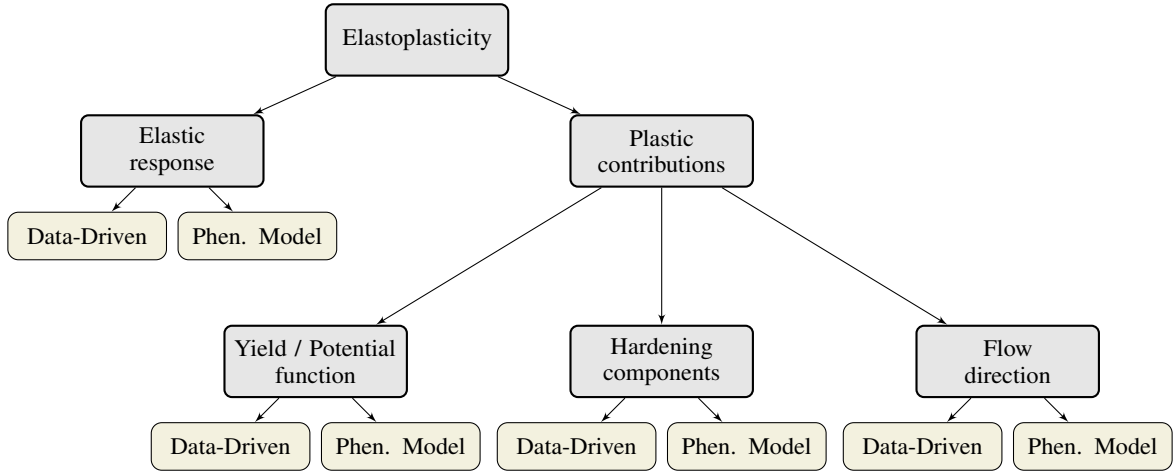
\begin{figure}
\begin{center}
\begin{tikzpicture}[scale=1, every node/.style={transform shape},node distance = 2cm, auto]
\footnotesize
\tikzstyle{decision} = [diamond, draw, fill=blue!20, 
    text width=4.5em, text badly centered, node distance=3cm, inner sep=0pt]
\tikzstyle{block} = [rectangle, draw, thick, fill=gray!20, text width=7em, text centered, rounded corners=1mm, minimum height=3em]
    
\tikzstyle{block2} = [rectangle, draw, fill=olive!10, text width=6em, text centered, rounded corners, minimum height=2em]
    
\tikzstyle{line} = [draw, -latex']

    \node [block] (init) {Elastoplasticity};

    \node [block,below left = .5cm and .8cm of init] (elastic) {Elastic\\response};
    \node [block2,node distance = 1.6cm, below left of=elastic] (DataEl) {Data-Driven};
    \node [block2,node distance = 1.6cm, below right of=elastic] (ModelEl) {Phen. Model};
    
    \node [block,below right = 0.5cm and .8cm of init] (plastic) {Plastic \\ contributions};

        \node [block,node distance = 2.8cm, below of=plastic] (hardening) {Hardening\\ components};
            
        \node [block2,node distance = 1.6cm, below left of=hardening] (Datahard) {Data-Driven};
        \node [block2,node distance = 1.6cm, below right of=hardening] (Modelhard) {Phen. Model};

\node [block,node distance = 4.6cm, left of=hardening] (initYield) {Yield / Potential \\ function};

        \node [block2,node distance = 1.6cm, below left of=initYield] (DataYield) {Data-Driven};
        \node [block2,node distance = 1.6cm, below right of=initYield] (ModelYield) {Phen. Model};
        
\node [block,node distance = 4.6cm, right of=hardening] (initflow) {Flow\\ direction};

        \node [block2,node distance = 1.6cm, below left of=initflow] (DataFlow2) {Data-Driven};
        \node [block2,node distance = 1.6cm, below right of=initflow] (ModelFlow2) {Phen. Model};

    \path [line] (init) -- (elastic);
    \path [line] (init) -- (plastic);
    
    \path [line] (elastic) -- (DataEl);
    \path [line] (elastic) -- (ModelEl);

    \path [line] (plastic) -- (initYield);

    \path [line] (initYield) -- (ModelYield);
    \path [line] (initYield) -- (DataYield);

    \path [line] (plastic) -- (hardening);
    
    \path [line] (hardening) -- (Datahard);
    \path [line] (hardening) -- (Modelhard);

    \path [line] (plastic) -- (initflow);
    
    \path [line] (initflow) -- (DataFlow2);
    \path [line] (initflow) -- (ModelFlow2);

\end{tikzpicture}    
\end{center}
\caption{Modular elastoplastic material modeling. 
The initial yielding can be fully DD or can separately include the equivalent stress measure and the yield stress.
Hardening components can e.g. include the deformation resistance or some form of hardening moduli. See \cite{vlassis2022component} and  \cite{fuhg2023modular}.} \label{fig::ModElasto}
\end{figure}


\subsubsection{DD plasticity modeled after elastoplasticity}
Elastoplasticity, in contrast to hypoplasticity, is characterized by modularity due to its clear distinction between elastic and plastic constitutive responses, with  historically established assumptions for each of these components.
In classical modeling, elastoplasticity is based on a formulation of the elastic response, a yield function description, an assumption on the direction of the plastic flow, and phenomenological models for the hardening behavior.
DD elastoplastic modeling exploits this modularity by choosing each of these components (or subcomponents) to be either represented by a DD formulation or a classical phenomenological model.
The general framework is summarized in Figure \ref{fig::ModElasto}.
In contrast to the DD hypoplastic models, described earlier, these models have both advantages and disadvantages. 
Some of the advantages include
\begin{itemize}
    \item More interpretability. 
    The input-output mapping of DD models is potentially easier to understand since each model is used as a representation scheme for a specific subproblem of the elastoplastic formulation. 
    \item Less data is needed. DD submodels can be chosen in areas where more data are available, while phenomenological models can represent the remaining components. This allows to reduce the reliance on big data. 
    \item Shift to one-to-one mappings. DD hypoplastic models intrinsically aim to find a consistent mapping for the whole path-dependent stress evolution. This leads to a one-to-many mapping problem which is solved e.g. with RNNs.
    On the other hand, DD elastoplastic models can only use ML models for components that can be modeled by simple one-to-one mappings, like yield function representations which can be decoupled from the time integration. Of course, this requires the access to data for these specific subproblems.
    \item Simpler enforcement of thermodynamic consistency.
    Since the majority of classical elastoplastic models were designed to be thermodynamically consistent, replacing single components of these models with specialized DD solutions allows for easier enforcement of physical constraints compared to ML models for plasticity modeled after hypoplasticity.
\end{itemize}
On the other hand, this modeling framework also has a disadvantage:
\begin{itemize}
    \item More constraints and assumptions. DD hypoplastic models as presented before are based on a very limited range of assumptions (such as the choice of the internal variables) and are therefore models that can discover the material behavior that underlies the training data.
    On the other hand, elastoplastic modeling needs to make more assumptions (e.g. the split between elastic and plastic components) and is therefore more restrictive in its ability to discover unknown physical processes.
\end{itemize}

\textbf{Uninterpretable ML approaches.} A number of ML approaches have been proposed that are modeled after elastoplasticity.
In \cite{furukawa2004accurate} both back stress and drag stress are incrementally updated using independent NNs to represent complex combined hardening laws under uniaxial cyclic loading, while the elastic law and the yield function are represented by traditional formulations.
\cite{jones2018machine} proposes an ML plasticity model that relies on the representation theorem of scalar-valued functions. By exploiting the modularity assumption, the authors define an NN for the elastic response and a separate NN for the flow rule, both of which can be treated as one-to-one maps. 
This allows for physical concepts such as material frame indifference and symmetry conditions to be enforced implicitly.
Another example is given by \cite{wang2019cooperative}, where established phenomenological models of the elastic response, the yield function, the plastic flow direction, and the hardening modulus are combined to find the best fit of a dataset through RL. In
\cite{stoffel2019neural}, a hybrid formulation is proposed where the plastic flow and the back stress evolution are represented by one NN, while the yield function is given by a phenomenological model.
\cite{Settgast2019} assumes that the elastic response is given by a linear law, whereas a NN takes the equivalent stress as an input to obtain the updated stress of the plastic region.
This allows for a smooth transition between elastic and plastic behaviors.
In \cite{Settgast2020}, the linear elastic law and equivalent stress formulation of the yield function are phenomenologically assumed, while the deformation resistance and the dilation angle between the deviatoric and spherical parts of the normalized direction of flow are trained by two separate NNs.
\cite{heider2020so} also splits the ML model into elastic and plastic components; the elastic trial stress is obtained from an NN which, if non-admissible, is mapped back using an RNN formulation. An objective loss function that includes frame indifference concepts is used for training.
\cite{jang2021machine} replaces the general return mapping framework with an NN that updates the predicted trial stresses. 
In a series of recent papers
\cite{vlassis2021sobolev, vlassis2022component, vlassis2022geometric}, the elastic law and the yield function evolution are treated as separately trainable ML models. An NN-based yield function is trained using a level-set hardening framework whose evolution is dependent on an internal variable.
Other work concentrates on just finding a ML representation of (parameterized) initial yield functions \cite{hartmaier2020data,park2021multiscale,shoghi2022optimal,schmidt2022data,xiao2022geometric}, whereas the remaining components are given by phenomenological models.
\cite{fuhg2022machine} uses input convex NNs to enforce the ML yield function representation to be convex.
Hybrid frameworks that locally improve phenomenological yield surfaces with ML models are investigated in \cite{ibanez2019hybrid,fuhg:hal-03619186}, with the latter enforcing physical constraints.
Different subproblems of elastoplasticity may use NN representations, such as the deformation resistance \cite{tsoi1991application,hwu1996comparative,hodgson1999prediction,liu2000prediction,sun2010development,li2012comparative,bobbili2015prediction,li2019machine,yang2020exploring,shang2022machine,li2022counterexample} or the isotropic hardening modulus \cite{zhang2020using}.
Other works employ SVR \cite{desu2014support,peng2020coupling} in the same context.

\textbf{Interpretable ML approaches.} Lately, interest is growing in deducing symbolic material models from data using symbolic regression and sparse regression.
The benefit of symbolic regression and sparse regression is that the models are expressed by short mathematical expressions that only involve a limited number of terms.
This increases their computational efficiency, physical interpretability, communicability, and calibratability.
In the context of elastoplasticity, this idea is followed by \cite{versino2017data,bomarito_development_2021,park2021multiscale} who use symbolic regression and by
\cite{flaschel2022discovering} who apply the sparse-regression-based EUCLID framework to discover symbolic expressions for the yield function and the hardening behavior from reaction force and full-field displacement data, thus avoiding the use of labeled stress-strain data pairs. In a similar fashion, but bypassing the development of hand-crafted libraries,  \cite{fuhg2024extreme} sparsify physics-augmented neural networks that are developed in a modular fashion, to obtain compact expressions related to yield function but also to isotropic and kinematic hardening. In \cite{Xu2024}, EUCLID is extended to pressure-sensitive and non-associated plasticity.
In \cite{flaschel2023automated}, the method is further generalized to the framework of generalized standard materials, i.e., a general framework which encompasses elasticity, viscoelasticity, plasticity and viscoplasticity.

\textbf{Model-free approaches.} In section \ref{hypoplasticity}, we have introduced approaches that rely on the so-called model-free approach to model plastic behavior directly on data of a given dataset more akin to the hypoplastic modeling approach. A different model-free elastoplastic-like technique is proposed in
\cite{ciftci2021data}, where transition rules between elastic and plastic responses are introduced which map to different subsets of the data.


\subsubsection{Summary}
We conclude this section with the summary table in Figure \ref{fig:Categorization}, which attempts to categorize some of the reviewed approaches based on the amount of data they need in input and on the number and type of constraints they are designed to satisfy. In general, a decreasing resort to constitutive modeling constraints is expected to correlate to an increasing need for data.

\begin{figure}
    \centering
\def\spc{7pt} 
\setlist[itemize]{wide=0pt, leftmargin=*, itemsep=0pt, topsep=0pt, after=\vspace*{-\baselineskip}, rightmargin=-\leftmargini}
\setlength{\extrarowheight}{3pt}
\begin{tikzpicture}
    \node[inner sep=\spc] (t)
        {
\begin{tabularx}{0.85\textwidth}{|X|X|X|}
\hline
Constraints & ML-encoded quantity & Examples\\
\hline
\begin{itemize}
\item Explicit internal variables
\item Fixed model functional form
\item Fixed flow direction
\item Fixed hardening types
\end{itemize}
&
\begin{itemize}
 \item
Model parameters
\end{itemize}
&
\begin{itemize}
\item
Phenomenological modeling
 \item
NLK models \cite{chaboche1986time}
\end{itemize}
 \\
  \arrayrulecolor{black!30}\midrule
  \begin{itemize}
\item Explicit internal variables
\item Fixed flow direction
\item Fixed hardening types
\end{itemize}
&
\begin{itemize}
 \item
Functional forms
\end{itemize}
&
\begin{itemize}
    \item EUCLID \cite{flaschel_unsupervised_2021,flaschel2022discovering}
    \item MAP123 \cite{tang2020map123}
\end{itemize}
 \\
  \arrayrulecolor{black!30}\midrule
  \begin{itemize}
\item Explicit internal variables
\item Fixed flow direction
\end{itemize}
&
\begin{itemize}
 \item Functional forms
\item Hardening types
\end{itemize}
&
\begin{itemize}
\item TANN \cite{masi2021thermodynamics}
\item Vlassis et al. \cite{vlassis2021sobolev}
\item Jones et al.  \cite{jones2018machine}
\end{itemize}
 \\
  \arrayrulecolor{black!30}\midrule
  \begin{itemize}
\item Explicit internal variables
\end{itemize}
&
\begin{itemize}
 \item Functional forms
\item Stress evolution law
\end{itemize}
&
\begin{itemize}
\item EUCLID \cite{Xu2024}
\end{itemize}

 \\
   \arrayrulecolor{black!30}\midrule
  \begin{itemize}
\item Latent space\newline  internal variables
\item Fixed flow direction
\end{itemize}
&
\begin{itemize}
 \item Functional forms
\item Hardening types
\end{itemize}
&
\begin{itemize}
\item Vlassis et al. \cite{vlassis2022geometric}
\end{itemize}

 \\
    \arrayrulecolor{black!30}\midrule
  \begin{itemize}
  \item Thermodynamically \newline constrained flow
\item Latent space \newline  internal variables
\end{itemize}
&
\begin{itemize}
\item Stress evolution law
\end{itemize}
&
\begin{itemize}
    \item Jones et al. \cite{jones2021neural}
\end{itemize}

 \\
     \arrayrulecolor{black!30}\midrule
  \begin{itemize}
  \item No thermodynamic \newline constraints
\item Explicit internal variables
\end{itemize}
&
\begin{itemize}
\item Stress evolution law
\end{itemize}
&
\begin{itemize}
\item Furukawa et al. \cite{furukawa1998implicit} 
    \item PODFNN \cite{huang2020machine}
\end{itemize}

 \\
      \arrayrulecolor{black!30}\midrule
  \begin{itemize}
\item No thermodynamic \newline constraints
\item Latent space/ implicit \newline  internal variables
\end{itemize}
&
\begin{itemize}
\item Stress evolution law
\end{itemize}
&
\begin{itemize}
    \item Mozaffar et al. \cite{mozaffar2019deep}
    \item Abueidda et al. \cite{abueidda2021deep}
\end{itemize}
 \\
\bottomrule
\end{tabularx}
};
\draw [-{Stealth[slant=0]},line width=2pt] (-7.5,8.0)--(-7.5,-8.0) node[pos=1.0,below] {\Large{Data}};
\draw [-{Stealth[slant=0]},line width=2pt] (7.5,-8.0)--(7.5,8.0) node[pos=1.02,above] {\Large{Constraints}};

\end{tikzpicture}
    \caption{Quantitative categorization of DD modeling of (hypo- and elasto-) plastic material behavior. More available data correlates to less need for constitutive modeling constraints. }
    \label{fig:Categorization}
\end{figure}

\subsection{Viscoelasticity}
One of the traditionally most challenging problems for constitutive modeling is that of viscoelasticity. Several mechanisms can be involved in the micromechanical processes that govern the macroscopic viscoelastic response, and correspondingly many characteristic timescales may have to be considered. Linear viscoelastic models are used in the regime of small deformations and are often represented through a combination of linear springs and dashpots, such as in the Maxwell, Kelvin-Voigt, Zener, and Generalized Maxwell models. Their nonlinear counterparts can account for nonlinear elastic contributions (such as those encountered in the context of hyperelasticity) and also nonlinear evolution equations. In general, and similar to what was discussed in the context of elastoplasticity, viscoelasticity can be cast in term of external (or observable) and internal variables. We can express the update of the stress $\boldsymbol{\sigma}$ and the internal variables $\boldsymbol{\xi}$ at time $t_{n+1}$ as
\begin{equation}
\mathcal{M}: \{\boldsymbol{\varepsilon}_n,\boldsymbol{\xi}_n\} \mapsto \{\boldsymbol{\sigma}_{n+1},\,  \boldsymbol{\xi}_{n+1}\} 
\end{equation}
which, in contrast to plasticity, are not subject to any additional constraints. This makes the form of the equations at hand similar in structure to the problem that an RNN is designed to replicate. Namely, as the RNN works with sequential data, the hidden state vector $a_{n+1}$ at time $t_{n+1}$ depends on the value of the hidden state vector $a_{n}$ at time $t_{n}$, whereas the output vector $y_{n+1}$ also depends on $a_{n}$, indicative of the history dependence that is being captured. This natural similarity has led to a lot of ML-enabled constitutive modeling approaches in the context of viscoelasticity to follow this general approach. The bulk of the literature can be separated in two main categories, those that focus on parameter estimation for known viscoelastic constitutive models, and others that utilize different variants of NN architectures for problems of different complexity. We begin with the latter.

\textbf{Uninterpretable ML approaches.} In a rather simple setting, but aiming to obtain CLs that generalize, the authors in \cite{al2006prediction} utilize a simple NN architecture to obtain what they refer to as a neural constitutive model for nonlinear viscoelasticity. The framework is also designed around the question of data availability and it allows for learning of rather simplified scalar-valued CLs from 1D creep test data, focusing on the non-zero component of the stress tensor for the particular test. The network takes temperature and initial stress level as additional inputs, aiming for further flexibility of the predictions.
In a similar fashion, \cite{kim2008multi} consider elastic properties as additional inputs to the NN and aim to learn nanoindentation creep responses under fixed loading conditions; however, these creep responses cannot be considered as full CLs that could e.g. be used in a FEM simulation.  In the same year of \cite{al2006prediction} another paper \cite{jung2006neural}, leveraging expertise in NN-based rate-dependent and rate-independent CLs, constructs a complete framework for a rate-dependent NN-based CL for viscoelasticity. Here, the stress and the strain are decomposed in volumetric and deviatoric components and the NN takes as inputs the current value of the strain and the values of stress and strain at the previous increment. The approach is shown to work satisfactorily and is also implemented in an Abaqus UMAT  so that structural problems can be tested. In their training data generation, the authors discuss data augmentation based on symmetry to train tensor-valued maps, and also obtain training data from structural simulations.

Several papers focus on learning the response of 1D experiments with ML approaches in the context of viscoelasticity, but in their essence, they are simple extensions of \cite{al2006prediction}. E.g., in \cite{kopal2017modeling} radial-basis-function NNs are utilized to learn the 1D dynamic response of thermoplastic elastomers, where the specific network architecture is chosen to accelerate the training process. More recently, \cite{jordan2020neural} focuses on the rate-dependent response of polypropylene; it utilizes simple NNs that take strain and temperature as input to learn the 1D large deformation response as a function of the loading strain rate. A robot-assisted testing system is deployed to generate large datasets, and Bayesian regularization is used to identify the network parameters. \cite{vu3822865physics} also focuses on learning the creep response, in this case, optical glasses, through an array of ML tools. The authors find these tools beneficial in terms of extrapolation to extreme temperatures compared to traditional phenomenological approaches. 

From a different starting point, but still focusing on the 1D case, \cite{basistov2018dynamic} addresses the fundamentals of history-dependent responses. Extending previous work where they had developed a model with associative (short-term) and hereditary (long-term) memory, inspired by a combination of the Kelvin-Voigt and the Maxwell models, the authors develop a new constitutive model with associative and hereditary memory as a system of integrodifferential equations. They then show that this system of equations can be approximated by a simple NN. They train based on 1D experimental data and test on non-monotonic 1D loading paths. The most interesting contribution of this work lies in the inherent interpretability of the approach since a direct connection is established between the NN architecture and the system of integrodifferential equations at hand. An early model proposed in \cite{oeser2009modeling} focuses on learning the time response of rheological material models stemming from a fractional differential equation using an RNN. A partial RNN is chosen, where the signal
flow occurs in the forward direction and the fading memory is realized by internal feedback connections. More recently, but in a similar fashion, \cite{chen_learning_2021} learns the solution of a specific boundary value problem with two non-zero stress components and encodes their response with an RNN. The work in \cite{graf2012structural} learns a fuzzy representation of the stress-strain response with an RNN recovering a fractional viscoelastic model, and \cite{freitag2013material} utilizes the RNN-based fuzzy CLs in a fuzzy-FEM setting to solve structural problems. A more recent work in the context of computational homogenization \cite{bhattacharya2022learning} establishes that the homogenized constitutive response may be approximated by a RNN. As an additional feature compared to previous approaches, a set of internal variables discovered in the learning of the homogenization procedure is tracked as a function of the history of the strain. \cite{upadhyay2023physics} extends the work of \cite{fuhg2022physics} to consider viscoelasticity (focusing on materials with limited memory) by employing tensor representation theorems for the deformation tensors and for the rate of these tensors. By performing the training based on limited experiments corresponding to specific deformation modes and by utilizing constrained GPR, the trained material laws are shown to proficiently generalize in the strain space and also with respect to the strain rate.


A multitude of works focus on ML or physics-informed (in the style of PINNs) solutions of PDEs involving the viscoelastic response of solids \cite{devries2017enabling,abueidda2022deep} but this is not the focus of the review here. We only mention \cite{xu2020inverse}, which deploys PINNs to train viscoelastic NN-based CLs based on limited sensor data by formulating a PDE-constrained optimization problem. The NN-based viscoelastic CL embedded in the structural optimization problem is not fundamentally different from the one in \cite{jung2006characterizing}. 

\textbf{Interpretable ML approaches.} For applications of symbolic regression in the context of viscoelasticity we refer to \cite{ratle_grammar-guided_2001} and more recently \cite{abdusalamov_discovering_2023}.
Further, the sparse-regression-based EUCLID framework, which utilizes a large library of phenomenological models and unlabeled data, is successfully used by \cite{marino2023automated} to identify linear viscoelastic CLs.
The same concept is extended in \cite{flaschel2023automated} to the theory of generalized standard materials, which naturally includes viscoelastic material behavior.

The approaches that focus on parameter estimation are not covered in depth in this review. In \cite{erchiqui2011neural}, from thin circular plate bubble experiments with temperature effects, and reducing the response to 1D, the material constants are learned for the Christensen viscoelastic model. \cite{hosseini2021optimized} aims to optimize the viscoelastic response through additives; based on experimental measurements, it calibrates the values of the loss and storage moduli utilizing several ML approaches. \cite{javidan2020experimental} fits the same parameters to compare to experimental data using a Kelvin-Voigt model. In the context of magnetorheological elastomers, \cite{saharuddin2020constitutive} suggests the use of NNs and Extreme Learning Machines (feed-forward NNs with a single hidden layer) to map shear strain and magnetic field to storage and loss moduli. \cite{Wang2021} uses GPR to fit the constants of a complex viscoelastic model based on experimental data; even though the task is seemingly straightforward, uniqueness is not guaranteed as the test data are only uniaxial and monotonic.

\textbf{Model-free approaches.} In \cite{eggersmann2019model}, the model-free approach proposed for plasticity is also adapted to viscoelasticity by relying on a differential representation of the material evolution history.  The model is tested within a monodimensional state space and applied to the analysis of truss structures.

\subsection{Damage and fracture}

The loss of integrity of a structural component can be modeled using different approaches depending on the process responsible for the material degradation. In the following, we distinguish between (i) diffuse damage, (ii) fracture, and (iii) fatigue. Damage describes a gradual deterioration of the material stiffness and/or strength without significant residual deformations. Fracture entails the formation and propagation of a crack, i.e. a discontinuity, which evolves following Irwin's or Griffith's criteria or extensions thereof. Note that the evolution of a crack in a continuous body can also be described using continuum damage models that allow for the localization of the damage parameter in bands with a limited but not vanishing thickness. 
Fatigue consists of the nucleation and propagation of a crack at subcritical load levels under repeated loading.

\subsubsection{Diffuse damage} \label{sct:dmg_mech}

The continuum damage mechanics approach pioneered by Kachanov \cite{Kachanov1958} aims at describing the gradual deterioration of the structural integrity of a material point when subjected to some type of action (e.g., displacements, forces, temperature changes, or aging). The main idea behind this theory is that, at the macroscopic scale, the reduction of stiffness and strength related to the material deterioration can be condensed in a scalar or tensorial internal damage variable. Often the onset of damage takes place after an initial elastic regime, which is followed by a stress-softening phase leading to an elasto-damage constitutive behavior. Since it cannot be directly measured, the damage variable belongs to the category of internal variables and, in the most common case, is assumed to be a scalar. Therefore, the related CL can be written as

\begin{equation}
\mathcal{M}_{\boldsymbol{\sigma}}: \{\boldsymbol{\varepsilon},\,{d}\} \mapsto \boldsymbol{\sigma}\,,
\label{eq:CL_damage}
\end{equation}

\noindent where the damage variable ${d}$ is governed by an often a priori postulated evolution law of the type

\begin{equation}
\mathcal{M}_{\dot d}: \{\boldsymbol{\varepsilon},\,\boldsymbol{q}\} \mapsto \dot{{d}}\,,\quad\text{subjected to}\quad \dot d\ge0\,,
\label{eq:evol_D}
\end{equation}

\noindent Here $\boldsymbol{q}$ is a vector collecting the set of mechanical and internal quantities governing damage evolution, while the non-negativity constraint is assumed to fulfill the second law of thermodynamics and takes the name of \textit{irreversibility condition}. This condition is the major responsible for the history dependence of the constitutive models accounting for material damage and it constitutes one of the main difficulties in the definition of a proper DD approach. On the other hand, it is of primary importance since it allows for a physically sound description of the unloading/reloading branches in the material response. 

In the following, we first provide an overview of the most relevant DD approaches including identification problems and constitutive modeling using NNs, and model-free approaches.
Most studies deal with NNs where damage is often implicitly accounted for while defining the CL and not explicitly introduced as an internal variable. No interpretable approaches are yet available.

\textbf{Uninterpretable ML approaches.} The earliest ML approaches in continuum damage modeling are devoted to parameter identification of available models. Although this review does not focus specifically on them, we propose here a brief overview due to their relevance in developing DD approaches for damage. \cite{Abendroth2003} identifies the parameters of the Gurson-Tvergaard-Needelman (GTN) model using a NN, which is trained using a set of load-displacement curves from FEM computations. The trained NN receives as input the experimental load-displacement curves and outputs the material parameters, thus surrogating the solution of an inverse problem. Aware of the poor extrapolation capabilities of the NN, the authors include in the cost function a penalty term to avoid extrapolation outside the range of the training data. Also, they point out that a reliable identification of the full set of GTN model parameters is not possible, hence, they restrict identification to a subset of the total parameters that need to be selected a priori depending on the material at hand. To extend the number of identified parameters, in \cite{Abendroth2006} the same authors propose a different identification method involving an NN to surrogate the solution of the forward boundary value problem for a small punch test. The NN is trained using FEM computations; it receives as input the displacement and the GTN parameters and it outputs the applied force. The parameter identification for a new set of experimental load-displacement data is then performed by adopting a successive quadratic programming algorithm. Similar identification approaches are adopted with minor modifications in other studies, e.g. \cite{Abbassi2013}.

The second class of approaches deals with the definition of elasto-damaging CLs using different NN architectures. \citep{Unger2009} surrogates the cohesive traction-separation law at the interface between concrete and steel using a NN trained through a set of FEM simulations. The NN takes as input the components of the interface separation vector and outputs the traction components. Hence, damage is not explicitly introduced but only implicitly accounted for. Since the NN is not informed by any physical requirement, the authors enforce some basic features of the CL by partly substituting ad-hoc calibrated linear relationships to the NN predictions. These include a vanishing traction value for vanishing or very large separation values, i.e. close to the unloaded state and to complete decohesion. Also, the obtained model is not able to distinguish unloading/reloading states and is thus limited to locally monotonic loading histories. In \cite{Unger2008} the same authors propose a ML constitutive model for the concrete bulk material to be used in a multiscale FEM framework. The approach involves the definition of a NN trained with micro-scale FEM analyses to surrogate the stress-strain law of the material, while an SVR algorithm with an exponential kernel is used to detect the unloading/reloading branches. The algorithm is formulated according to a criterion similar to a limit surface whose extension depends on a history variable defined as the maximum strain reached during the loading history of a point. To exclude any mesh dependence related to the underlying (implicit) local damage approach, the size of the finite elements at the macro-scale is used as an input in the NN. In turn, this calls for a training set that includes data for different sizes of the domain at the micro-scale. Although the formulation of the approach is general, the authors illustrate its performance only for 1D cases. 

An ML-enhanced multiscale framework is also proposed by \cite{Settgast2019,Settgast2020} to model elastoplastic damaging foams. In \cite{Settgast2019}, the authors surrogate the macroscopic stress-strain relationship for proportional loading with a NN trained using lower-scale FEM computations. The homogenized local tangent stiffness tensor is computed directly by derivation of the NN mapping and the cost function is complemented with a penalty term to prevent large differences in the order of magnitude of the neural weights. A different approach is explored in \cite{Settgast2020}, where the definitions of homogenized limit surfaces, flow directions and stiffness deterioration due to the (implicit) damage are ascribed to three different NNs. In particular, the definition of a limit surface allows to distinguish between the dissipative and elastic unloading/reloading stages.

A multiscale framework to define the behavior of a damaging poroelastic material is presented in \cite{Wang2018a}, where three scales (i.e., micro-, meso- and macro-) are accounted for. The authors use LSTM-type RNNs to upscale the material parameters between micro- and mesoscale and between meso- and macro-scale. The NNs are trained using discrete element and finite element computations for the micro- and mesoscale respectively, while the analyses at the macro-scale are performed using the FEM. The material models at each scale are defined using directed graphs where the relations between the various parameters involved are known as physical or empirical relationships. Also in this case the damage parameter is not explicitly defined but is implicitly considered in the CL. The adoption of RNNs makes the model intrinsically history-dependent, thus allowing for easy discrimination between different states, such as unloading/reloading branches.  Another addressed point is the objectivity of the material response given by the RNN, which is not satisfied a priori. The authors propose a method to achieve material objectivity based on a spectral representation of the training data, which effectively reduces the deviations in stresses and energy between different observer frames when evaluating the same system.

Two different NNs are adopted in \cite{Yan2020} to define the macroscopic CL of a fiber/matrix composite material. The first NN is used to surrogate the stress-strain relationship of the material; also in this case the damage variable is not explicitly defined and the model does not include any unloading/reloading criterion.  The second network is used to identify if the damage takes place in the fibers or in the matrix without explicitly accounting for them in the simulation. 

\cite{Fernandez2020} uses NNs to surrogate the traction-separation law between adjacent grain boundaries at different temperatures. In particular, they adopt a standard NN to define the interface secant stiffness, which is then multiplied by the separation vector giving a ResNet-like CNN \cite{he2016}. Differently from an RNN, the proposed architecture is unable to process temporal information and, hence, it can be used only for monotonic loading history. On the other hand, the major advantage of this approach compared to an RNN is that it drastically limits the amount of data needed for training. The authors illustrate the approach in 2D training of the NN by using a dataset of molecular dynamics simulations. Also, a procedure to optimize the NN architecture is proposed. 

PINNs are adopted by Haghighat et al. \cite{Haghighat2022}
to surrogate a coupled damage/plasticity model. The introduction of physical and modeling constraints allows us to include in the PINN conditions such as damage irreversibility, vanishing stress for vanishing strain, and complete failure after a critical damage value, and to automatically detect elastic loading/unloading/reloading states. However, the approach requires a very large amount of training data to achieve reasonable accuracy.

Among the approaches that do not belong to the aforementioned categories, \cite{wang2019meta} proposes a supervised CL discovery approach exploiting directed graph theory to automatically generate different models. The authors test the proposed approach to obtain the traction-separation law of a cohesive interface with implicit damage for 2D problems. The approach requires the definition of the parameters that possibly govern the CL and a set of \textit{rules} that the relationships between the various parameters must satisfy to create a valid CL. The generation of a model takes the name of \textit{game of model generation} and each \textit{move} during a game connects two parameters with a relationship until the input parameters are connected with the outputs. The completed model is then evaluated by assigning a \textit{score} based on the comparison between the model predictions and a set of material observations. Unlike in \cite{Wang2018a}, here the relationships between the parameters are encoded by an RNN, which is trained using a set of discrete element analyses. A deep RL algorithm is used to iteratively improve the models game after game, based on the evaluation of the probability that a certain move (i.e., a connection between two parameters) is selected by a \textit{gamer} at a certain point of the game along with its expected contribution to the final score. Probabilities of moves and expected score contributions are predicted using an NN enhanced by a Monte Carlo tree search algorithm. A new move is then selected so as to maximize the final model score. The major feature of this supervised approach is that, once the game and a dataset of material observations are defined, no operator intervention is needed since the algorithm is able to learn how to improve its predictive capabilities from the previous games played. Also, the adoption of an RNN ensures that the model history dependence is included.

\textbf{Model-free approaches.} A model-free approach is adopted in \cite{Karapiperis2021} to define the macroscopic behavior of granular materials subjected to different loading conditions. The material dataset, generated with the level-set discrete element method, includes also a parametrization that allows to describe the history dependence of the material behavior. The definition of the latter is critical to obtain meaningful results and the authors propose to use either the dissipated energy or a set of internal variables known to satisfactorily describe the microstructural arrangement of a granular material. Both approaches give similar results but the method to obtain the material dataset must be selected with particular care since it must allow for the definition of the desired quantities. 

\subsubsection{Fracture}

The propagation of a crack inside a solid can be described by means of different fracture mechanics models depending on whether the material exhibits a linear or nonlinear elastic or elastoplastic behavior. Here we will focus only on linear-elastic fracture mechanics approaches since, to date, the available literature about DD methods deals only with this class of behavior. Hence, the available theories revolve around Irwin's or Griffith's criteria and extensions thereof. Hence, they call for the definition of a stress intensity factor (SIF) or of an energy release rate that are functions of the current size of the crack $a$. Also in this case the fulfillment of the second law of thermodynamics calls for the introduction of an irreversibility condition on the crack size, namely $\dot{a}\ge0$. 

\textbf{Uninterpretable ML approaches.} One of the earliest approaches is reported in \cite{Theocaris1993, Panagiotopoulos1999}. Here the authors use an NN to solve problems involving a crack contained in a linear elastic domain, considering or not the unilateral constraint given by crack face contact. The definition of the weights and of the activation functions is inspired by the variational principle of energy minimization; the outputs are taken as the displacements, the weights carry the information related to the elastic properties of the material and the cost function is the elastic energy of the system. Thus, this can be considered a prototype of the physics-informed DD approach. The approach does not account for any propagation criterion and is used to solve both direct and inverse problems. In forward problems, the kinematic and static fields are computed for a given external load, while the weights of the NN are defined using FEM computations. The authors observe a faster convergence of the constrained problem compared to its unconstrained counterpart. For the constrained case, the computational cost is lower than with FEM with singular elements, while for the unconstrained case, the opposite is true. In inverse (identification) problems, a set of displacements is supplied to the algorithm and the elastic properties of the material (i.e. the weights) are obtained following a backpropagation solution scheme. 

Different contributions deal with the computation of the SIF for different geometries and boundary conditions (see, e.g., \cite{Liu2021, Wang2021}). In particular, \cite{Liu2021} proposes a NN to compute the SIF for a micro-cantilever beam. Training is performed using a set of FEM computations that is adaptively enriched through the targeted addition of sampling points within the portion of the state space where less accuracy is expected. The accuracy is estimated pointwise as the maximum deviation within thousands of trained NNs and their average prediction. Although the method exploits the linearity between load and SIF to achieve a more general result, the results depend on the geometry considered during training. The same applies to  \cite{Wang2021}, where the critical SIF is obtained for a Brazilian test. Here the authors also explore the possibility of using ML approaches different than NNs, such as decision trees, random forest regression, and extra regression trees, concluding that the latter two approaches are more efficient than NNs.

An effective method to describe crack nucleation and propagation using a damage parameter that localizes in crack-like narrow bands is the phase-field approach to fracture \cite{Bourdin2000}. 
In this context, ML-related work has concentrated on the solution of the governing PDEs obtained from the minimization of the phase-field energy functional. 
\cite{Aldakheel2021} uses NNs to surrogate the solution of these PDEs. 
Two different NNs are trained, one for the linear elastic bulk material and the second for the evolution of the damage variable, and they are embedded into a FEM code either together or alone. 
The solution of phase-field fracture problems with an approach known as the deep Ritz method (DRM)~\cite{yu2018deep} is pursued in some recent studies \cite{Motlagh2022, goswami2020transfer, goswami2020adaptive, Manav2024}. Unlike PINNs, for problems in which the governing PDEs stem from the minimization of an energy functional, the DRM directly minimizes the energy functional instead of the PDE residual. The approach in \cite{Manav2024} solves examples of crack initiation, propagation, kinking, branching, and coalescence within one single numerical setup.
An operator learning approach~\cite{lu2021learning}, namely variational DeepONets, has also been applied to predict the crack path in a problem involving crack propagation in quasi-brittle materials~\cite{goswami2022physics}. The potential of all these approaches is not to be applied to the solution of a single boundary value problem, but rather to learn solutions to parametric phase-field fracture models. In this setting, NNs can be trained on a few realizations of the parameter space and results can be inferred online for all other realizations, leveraging the true potential of the DRM and of neural operators. 

An SVR approach based on a novel kernel function is proposed in \cite{Feng2021} to surrogate the solution of a phase-field problem accounting for the stochastic distribution of the material and geometry parameters. The pool of input parameters includes the fracture and elastic properties of the material and the point of application of the load, while the geometry is considered deterministic. The SVR training is performed through Monte Carlo sampling, using FEM results where the material and geometry parameters are sampled from a given statistical distribution and is enhanced through a clustering technique. 

\textbf{Model-free approaches.} 
A model-free approach based on variational principles is proposed in \cite{Carrara2020}, where the fracture-related constitutive behavior is encoded into a dataset of material observations including crack position $a$ and dissipated energy per unit surface (i.e., the fracture toughness). The constraint set for the admissible state space involves conditions encoding either a global or a local minimization of the total energy of the system subjected to a crack irreversibility condition. The authors propose and discuss different metrics defining the generalized distance and investigate the effect of noise in the material dataset on the final results. In \cite{Carrara2021} the approach is extended to rate-dependent fracture mechanics. The authors show that some classical constitutive assumptions, such as irreversibility or monotonicity, are redundant in a model-free setting since they are implicitly encoded in the material dataset. Note that both cases are defined and tested for setups for which an analytical expression of the energy release rate is available.

\subsubsection{Fatigue}

Classic linear elastic fracture mechanics states that crack propagation is triggered if a relevant quantity (either the SIF or the energy release rate) reaches a certain critical value. However, the experimental evidence demonstrates that, even below the critical value, a crack can still propagate but to an extent that becomes measurable only after the application of several load cycles, giving rise to fatigue crack propagation. 

To date, fatigue is still typically described using empirical laws calibrated on wide experimental datasets, i.e. in an inherently DD fashion. The most prominent mathematical description of the fatigue behavior that can be interpreted as a CL is due to Paris \cite{Paris1963} and reads 

\begin{equation}
\frac{da}{dN}=C\Delta K^m\,,
\label{eq:fatigue}
\end{equation}

\noindent where $N$ is the cycle number, $da/dN$ is the fatigue crack growth rate, $\Delta K$ is the SIF range spanned by the crack tip in a single load cycle, and $C$ and $m$ are two material parameters. 

\textbf{Uninterpretable ML approaches.} In \cite{Lee2005} the authors use an NN to obtain the fatigue crack growth rate and the remaining fatigue life given the applied load, the crack size, and additional information commonly used to estimate the fatigue life. This includes a pre-calibrated empirical function that should mimic the role of $\Delta K$ and some easily accessible experimental parameters, making this method more useful for diagnostics than for modeling. 

\cite{Rovinelli2018} adopts a Bayesian network approach to predict the probability of activation of a certain slip plane and the associated fatigue crack growth rate in body-centric cubic polycrystal materials. The network is trained using crystal plasticity computations and experimental data. 

\textbf{Model-free approaches.} \cite{Carrara2021} illustrates that the model-free approach for rate-dependent fracture mechanics can be conveniently adapted to fatigue if the time variable is substituted by the number of cycles $N$. In this case, the dataset of material observations is composed of crack growth rate-driving force pairs, which replace (\ref{eq:fatigue}). This approach is able to automatically account for characteristic features of the fatigue behavior, such as the presence of a threshold for $\Delta K$ below which no fatigue effects are triggered.

\subsection{Multiphysics}


Many processes of engineering interest involve multiple physical mechanisms coupled with the mechanical behavior. 
In multiphysics CLs, mechanical quantities such as strains and stresses are insufficient to describe the material state; additional quantities related to the e.g. thermal, hydraulic, or chemical responses may need to be additionally considered. 
The complexity of the experiments involving multiphysics processes makes the identification of the related CLs even more challenging than in solid mechanics. 
More specifically, identifying appropriate material descriptors and establishing their functional dependence on the material state in the multiphysics process while satisfying the laws of thermodynamics and accurately modeling experimental data is a complex endeavor. ML algorithms may offer a promising approach to accelerate the discovery or surrogation of material models in such scenarios. To limit the scope of the following review to a manageable extent, we focus here on multiphysics problems relevant to geomechanics and do not cover multiphysics CLs arising e.g. in thermo-, electro-, or magnetomechanics. Examples of using DD models for the latter can be found in \cite{kalina2024neural,klein2024nonlinear}. 


The strain-stress state of the material in a multiphysics setting may be influenced by a variety of material descriptors, such as the void ratio in geomaterials. The behavior of these descriptors, as well as the strain and stress states themselves, may be influenced by other unknown fields, such as fluid pressure. Moreover, these material descriptors may interact with each other, and their causal relationships are not always clear. To review recent developments in multiphysics constitutive modeling, we can classify problems into three types: (1) problems in which the material descriptors and their causal relationships are known, but the functional form of the CL is unknown; (2) problems in which the material descriptors are known, but their causal relationships and functional form are unknown; and (3) problems in which the material descriptors, their causal relationships, and their functional form are all unknown.


In the first category, traditionally, models are handcrafted based on experimental or simulation data. 
Advances in computational multiscale methods such as $\text{FE}^2$ 
\cite{feyel2003multilevel,kanoute2009multiscale} do not need any functional forms for the constitutive model, and they can be built based on first principles of thermodynamics to automatically avoid physical violations. Although these computational approaches are accurate, they usually suffer from high computational cost already in the mechanical context, and even more so in a multiphysics setting. This is because not only the iterations between two scales are required but also different physics dictate distinct temporal and/or spatial resolutions. 
In this regard, finding a functional form may seem to be a more computationally efficient approach, which additionally enjoys the advantages connected to interpretability. 

\subsubsection{Methods to define the functional form of the CL}

\textbf{Uninterpretable ML approaches.} \cite{dehghani2020poroelastic} employs fully connected NNs to model macro-scale parameters for an isotropic poroelastic medium, without enforcing physics constraints in their surrogate modeling approach. In a related work \cite{dehghani2021ann}, a similar framework is employed for the anisotropic finite-strain regime, where microstructure descriptors such as porosity and Poisson's ratio are used to learn macroscopic homogenized values. This approach bypasses the need for computationally expensive fine-scale simulations in concurrent multiscale frameworks. \cite{abueidda2021deep} develops a surrogate model using GRU and temporal CNNs, which predicts the homogenized stress tensor and the temperature based on micro-scale specimen information and loading conditions. Their approach does not take into account thermodynamics rules. \cite{ashworth2022machine} utilizes surrogate constitutive models to simulate mass transfer in dual-porosity materials at the macro-scale, reducing the computational cost in classical multiscale simulations. They demonstrate the effectiveness of autoregressive NNs, which have the advantage of being easier to implement and train compared to RNNs. \cite{heider2021offline} investigates the use of surrogate modeling for unsaturated soils; complex retention curves with different wet and dry branches and anisotropic permeability models are determined in a purely DD manner using fine-scale simulations. Additionally, the study proposes an automated framework based on RL to identify the best set of hyperparameters.
Overall, surrogate modeling approaches based on  NNs or other types of differentiable approximators (e.g., Gaussian processes) have the advantage of being scalable for high-dimensional problems with different types of algebraic and ODE constraints. 

\textbf{Interpretable ML approaches.} In multiphysics CLs, the material behavior can be expressed by algebraic equations, in simple cases, or in general, by a system of coupled high-dimensional ODEs or PDEs. Although symbolic regression methods find the explicit form of these equations, they become more and more computationally inefficient when the number of material descriptors grows, or the desired functional form is not algebraic, i.e., it is given by ODEs or PDEs \cite{de2018greedy}. Moreover, incorporating physics (e.g. thermodynamic) constraints in their discrete search algorithm may further increase the computational cost. Recently, methods based on RL have been developed to improve the computational burden of the symbolic regression search space \cite{petersen2019deep}. 
Based on the current challenges regarding symbolic regression, sparse regression appears as a promising option for interpretable ML approaches; surprisingly, to the best of our knowledge, it has received very little attention so far. A first approach in this direction has been suggested by \cite{fuhg2024polyconvex}, who propose a polyconvex neural network-based thermoelastic framework that achieves interpretability by sparsifying the number of parameters in large neural networks down to only a few remaining ones.

\textbf{Model-free approaches.} 
\cite{bahmani2021kd} extends the non-parametric distance-minimization paradigm (i.e. the model-free approach) to poroelasticity problems. It is assumed that the material descriptors for a coupled poroelastic problem are known (based on the effective stress principle and Darcy's law), but no functional forms are assumed for the relations between strain, effective stress, pore pressure, and Darcy's velocity. This approach minimizes assumptions on the modeling part.
However, the model-free approach is known to be data-hungry \cite{he2020physics, bahmani2022manifold}, and this feature becomes even more prominent in a multiphysics setting. To address this issue, the authors introduce variationally consistent multi-fidelity formulations where a model-based method for one field is hybridized by an entirely DD assumption for the other field based on data availability and quality. Due to the discrete nature of the distance-minimization problem, the brute-force approach for searching in the data (at inference) becomes inefficient for a multiphysics problem. They develop an efficient data structure based on kd-trees to reduce the computational burden exponentially.

\subsubsection{Methods to define the causal relationship between descriptors and functional form of the CL}

In this category, the problem is more challenging since not only the functional form is unknown, but also it is not known how the material descriptors impact each other. The causality between material descriptors may help to find more generalizable functional forms. More importantly, it helps explain the underlying physical process, which is crucial for a mechanics problem involving e.g. failure or fracture. \cite{wang2019meta} formulates such a problem as a directed acyclic graph (DAG) problem where the nodes are material descriptors, and the goal is to find edges that lead to an information flow best describing the input-output relationships. In this work, each edge uses an RNN model to incorporate any possible path-dependence which is common in plasticity and visco-elasticity problems. The approach does not include thermodynamics constraints which could be addressed by recent ideas from the PINN community \cite{raissi2019physics,masi2021thermodynamics,vlassis2021sobolev}. Moreover, it is assumed that the DAG structure remains constant, which may not apply to other path-dependent problems. 

\subsubsection{Methods to define the material descriptors, their causal relationships, and the functional form of the CL}

The complexity further increases in the third category since the set of plausible material descriptors is not known a priori. Historically, mechanicians have encountered several such issues; e.g., the existence of the fabric tensor as an additional material descriptor for describing the plasticity of granular materials was not apparent based on classical theories and assumptions \cite{dafalias2004simple}. Therefore, it could be beneficial to design an intelligent system to find the optimal material descriptors based on the material state information. \cite{sun2022data} proposes a probabilistic framework to find the plausible set of material descriptors with their associated confidence interval. They provide a systematic way to incorporate uncertainty in causal discovery and predictions. They study a granular system and show that their framework can recover some classical theories regarding the fabric tensor. One limitation of their work is the static assumption on the DAG structure which could be subjected to change at different loading states.


\section{Evaluation, verification, validation and their challenges}

\subsection{Performance metrics}
To evaluate the performance of different DD approaches, a set of unbiased performance metrics is necessary. A major problem in DD for mechanics compared to other more established applications is the lack of standardized tests to evaluate performance and even an agreement on metrics to be used for that evaluation. Here, even though our goal is not to establish or solidify such consent, we try to guide the development of a framework around this task. To this end, we propose the following metrics: 
\begin{itemize}
    \item \textbf{Accuracy} -- the quality of the predictions, within the training data, defined by a metric that measures discrepancy. This error is generally established by sampling errors between ground-truth data and predictions.
    \item \textbf{Precision} -- the sensitivity of the predictions with respect to the amount of available data and their noise, the sampling method adopted and the distance between data points in the training set. 
	\item \textbf{Physical/mechanical consistency} -- the ability to fulfill basic physical/mechanical laws or principles with a general validity, namely not related to specific constitutive modeling choices. Examples are objectivity, material stability, energy balance, vanishing energy and stress associated with an undeformed state, and positive variation of the energy for increasing external loads.
    \item \textbf{Interpretability} -- the amount of information that can be understood by modelers and decision makers from a calibrated DD constitutive model.
    \item \textbf{Generalization ability} -- the ability of the approach to provide accurate predictions outside of the training set. This has as a prerequisite the efficient interpolation on unseen training data, but more importantly, it evaluates the accuracy of the model in extrapolation outside the training domain.  
    \item \textbf{Cost/performance trade-off} -- the time and computational cost required to complete the workflow for predictions. This includes the time and computational cost required to train, validate, test, and verify the models, the time required to run the code as well as the demand for data to ensure the required level of accuracy, precision, and generalization. 
    \item \textbf{Robustness} -- the sensitivity of the above list of metrics with respect to the amount of available data, noise in the data, the distance between the predicted loading path and the calibration data, and the repeatability and reproducibility of the performance metrics. 
     \item \textbf{Stability} -- the sensitivity of the above list of metrics to perturbations of the free parameters of the DD approach.     
\end{itemize}

The information provided by the proposed metrics helps to evaluate also some specific characteristics of the obtained DD CL in relation to the required employment. A non-comprehensive list is provided as follows
\begin{itemize}
\item \textbf{Fidelity} -- the level of accuracy, precision, and generalization achieved in performing a given task (e.g., optimize a component, predict the behavior of a material under given conditions, better understand the physics behind a studied phenomenon). Depending on the applications it can encompass or not the evaluation of the physical/mechanical consistency.
\item \textbf{Confidence} -- the precision that can be attained in reproducing the phenomenon of interest with a certain accuracy given a set of calibration data. This metric essentially quantifies the uncertainty related to the obtained CL. 
\item \textbf{Data hunger} -- the amount of training data needed to reach the desired level of fidelity. 
\item \textbf{Numerical behavior} -- possibility to integrate the obtained model in a given numerical framework, its associated cost/performance ratio, and convergence with respect to the spatial and, if relevant, temporal discretization. 
\item \textbf{Reliability/reproducibility} -- the sensitivity of the metrics with respect to the source of the calibration data (e.g., different codes for the simulations or laboratories for the physical tests) and of the training code adopted. 
\item \textbf{Trustworthiness} -- a combination of all the metrics relevant to the task at hand that give a synthetic global score of the performances of the DD CL.
\end{itemize}

Whenever adopting a DD approach to reproduce a CL it should be good practice to define which important metrics apply to the case/application at hand and how the operator intends to objectively measure them. Moreover, the definition of the performance metrics should assist in comparing either different approaches to model the same or similar phenomena (e.g., uninterpretable vs. interpretable approaches or comparison between different NNs) or the same approach but on different set of training data (e.g., same material but different experiments, or different materials but same phenomenon of interest). Another field of application of the performance metrics is the validation and verification task. In particular, the definition of the metrics should provide an unbiased and objective way to assess whether the obtained CL is able to reproduce the behavior under investigation and how much the obtained results can be trusted. This topic is better analyzed in the following section.


\subsection{Verification and validation}

Verification and validation (V\&V) is necessary for a confident quantification of engineering predictions, especially in the context of making decisions in high-consequence scenarios.
Generally, in the computational solid mechanics community verification addresses the process of confirming that a mathematical model is correctly represented by its governing equations, appropriately discretized in a numerical setting, and that the solution complies with relevant accuracy requirements \cite{schwer2007overview}. Conversely, validation refers to quantifying the degree of accuracy to which a model represents the investigated phenomenon \cite{sargent2009verification}. In the following, we broadly ignore V\&V from a software development perspective (we refer to \cite{wallace1989software, thacker2004concepts} for more information) and assume that the code is error-free and well implemented. Also, we refrain from proposing or summarizing guidelines about how to verify numerical convergence and stability of a code, for which we refer to \cite{Zienkiewicz2005, Hughes2008}.

When compared to traditional phenomenological modeling, where the constitutive parameters of the material models are calibrated using experimental data, DD constitutive modeling involves an additional layer of difficulty.  In particular, the decision logic of a DD approach is oftentimes opaque even to their designers, therefore a careful, separate, and independent V\&V process to ensure they are reliable and safe to use must be performed.

\begin{figure}[!hbt]
\footnotesize
\tikzstyle{rect} = [rectangle, rounded corners, minimum width=3.5cm, minimum height=2em,text centered, draw=black, fill=gray!20]
\tikzstyle{rect2} = [rectangle, rounded corners, minimum width=2.5cm, minimum height=2em,text centered, draw=black, fill=green!5]

\tikzstyle{rectInv} = [rectangle, rounded corners, minimum width=3.5cm, minimum height=1cm, draw=blue!0, fill=blue!0]

\tikzstyle{rectInvSm} = [rectangle, rounded corners, minimum width=1.5cm, minimum height=1cm,text centered, draw=blue!0, fill=blue!0]

\tikzstyle{dia} = [ ellipse, minimum width=4.2cm, minimum height=1cm,text centered, draw=black, fill=blue!20]

\tikzstyle{rectEmpy} = []
\tikzstyle{line} = [thick,draw, -latex']
\tikzstyle{line2} = [draw,thick]
\tikzstyle{arrow} = [thick,->,>=stealth]
\begin{center}
\begin{tikzpicture}[align=center, node distance=2cm]
\footnotesize

\node (proc) [rect] {PROCESS/PHENOMENON \\DEFINITION};

\node (interest) [rect, below right =.75cm and -1.5cm of proc] {Quantities of Interest - QoI \\ (Output)};
\node (variab) [rect, below left =.75cm and -1.5cm of proc] {Variables/parameters \\ (Input)};

\coordinate (belowvariab) at ($(variab)+(0cm,-.6cm)$);
\coordinate (belowinterest) at ($(interest)+(0cm,-.6cm)$);
\coordinate (centerintvar) at ($(proc)+(0cm,-2.1875cm)$);

\draw[line2](variab) -- (belowvariab) -- (belowinterest) -- (interest);

\coordinate (abovevariab) at ($(variab)+(0cm,.6cm)$);
\coordinate (aboveinterest) at ($(interest)+(0cm,.6cm)$);
\draw[line2](variab) -- (abovevariab) -- (aboveinterest) -- (interest);

\coordinate (abovecenterintvar) at ($(proc)+(0cm,-.98cm)$);
\draw[line2](proc) --  (abovecenterintvar);

\node (conceptModelT) [rect,  below = 3cm of proc] {Conceptual model};

\draw[arrow](centerintvar) --  node [pos=0.6,right] {Mechanistic assumptions\\ Physical concepts}  (conceptModelT);

\node (conceptModel) [rectEmpy,  below = 0.3cm of conceptModelT] {};

\node (verification) [rect, minimum height=2cm,  text depth =.5 cm,below right = 1.75cm and 2.3cm  of conceptModelT, fill=red!20] {\textbf{VERIFICATION} \\ {$\,$} \\\parbox{3.2cm}{- Sensitivity analysis \\ - Physical consistency \\ - ...}};

\node (veriftest) [dia,  below = .75cm of verification.south] {Verification accepted?};

\node (mathModel) [rect, below = 1cm of veriftest] {\textbf{CONSTITUTIVE LAW} \\ \textbf{(Mathematical model)}};

\node (Verif) [rectInv, below right = 0.1cm and -0.5cm of mathModel] {\textcolor{blue!70}{Code \&} \\ \textcolor{blue!70}{Calculation} \\ \textcolor{blue!70}{Verification}};

\node (equi) [rect, below left = -0.05cm and 0.5cm of mathModel] {\parbox{3cm}{\textcolor{blue!70}{- Governing equations} \\ \textcolor{blue!70}{- Boundary conditions} \\ \textcolor{blue!70}{- Geometry}}};
\node (experiment) [rect, below left = 2.5cm and 2.3cm  of conceptModelT] {Experiments design};
\node (Calexperiment) [rect2, minimum height=2cm, below left = 1.75cm and -1.3cm  of conceptModelT, text depth = 1.5 cm] {Calibration data:};
\node (input) [rect, minimum width=2cm, minimum height=.3cm, below =-1.4cm of Calexperiment] {Input};
\node (output) [rect, minimum width=2cm, minimum height=.3cm, below =-.7cm of Calexperiment] {Output};
\node (training) [rect2, minimum height=2cm, below left = 1.75cm and -4.8cm  of conceptModelT, text depth = 1.5 cm, fill=olive!10] {Training:};
\node (unsupervised) [rect, minimum width=2cm, minimum height=.3cm, below =-1.4cm of training] {Unsupervised};
\node (supervised) [rect, minimum width=2cm, minimum height=.3cm, below =-.7cm of training] {Supervised};

\node (appsel) [rect,below left= -3cm and -1.2cm of training] {DD method selection};

\node (Valexperiment) [rect, below = 1.5cm of experiment] {Validation tests};

\node (expResult) [rect, below = 2.0cm of Valexperiment] {Validation outcome \\ (Ground truth)};

\draw [arrow] (Valexperiment) -- node [pos=0.5,left] {UQ} (expResult);

\node (CompModel) [rect, below = 1.2cm of mathModel] {Digital twin};

\coordinate (leftdigital) at ($(CompModel.west)+(-5.77cm,0.0cm)$);

\draw [arrow] (Calexperiment) -- (leftdigital) -- node [pos=0.5,below] {model-free approach}  (CompModel.west);

\node (result) [rect, below =1cm of CompModel] {Simulation outcome};

\node (Comp) [rect, minimum height=2cm,  text depth =.4cm,left =2.3cm of result, fill=red!20] {\textbf{VALIDATION} \\ {$\,$} \\\parbox{3.2cm}{- Fidelity \\ - Numerical behavior \\ - ...}};

\node (conc) [dia, below= 1.1cm of Comp] {Validation accepted?};

\node (Accept) [rectInv, right = 1.1cm of conc] {Model Validated \&\\Verified};
\draw [arrow] (conc) -- node [pos=0.2,below] {Yes}  (Accept);

\coordinate (leftVal) at ($(conc)+(-7.8cm,0.0cm)$);
\coordinate (upVal) at ($(conceptModelT)+(-7.8cm,0.0cm)$);
\draw[arrow]  (conc.west) --node [pos=0.07,below] {No} node [pos=0.6,above] {Revise conceptual or \\ DD model or \\ recalibrate CL } (leftVal) -- (upVal) -- (conceptModelT) ;

\draw [arrow] (conceptModel.center) -| node [pos=0.3,below] {Simulation/ Physical} (experiment);
\draw [arrow] (mathModel) -- (CompModel);

\draw [arrow] (CompModel) -- node [pos=0.5,left] {UQ} (result);

\draw [arrow] (result) -- (Comp.east);

\coordinate (belowExpres) at ($(expResult.south)+(0cm,-2.405cm)$);

\draw [arrow] (expResult) -- (belowExpres) -- (Comp.west);

\draw [arrow] (Comp) -- (conc);

\draw [arrow] (input) -- (unsupervised);
\draw [arrow] (output) -- (supervised);

\path[arrow]
(input.east) edge [out=0, in=-180]  (supervised.west);

\draw [arrow] (training) -- (verification);
\draw [arrow] (verification) -- (veriftest);

\coordinate (rightConc) at ($(conceptModelT)+(8.35cm,0.0cm)$);
\coordinate (rigthvaltest) at ($(conceptModelT)+(8.3cm,-5.335cm)$);
\coordinate (test) at ($(conceptModelT)+(8.3cm,-5.335cm)$);

\draw [arrow] (veriftest) -- node [pos=0.25,right] {Yes}  (mathModel);

\path let \p1=(veriftest.west), \p2=(Calexperiment.south) in coordinate (notverif) at (\x2+0.3cm,\y1);
\path let \p2=(Calexperiment.south) in coordinate (calverif) at (\x2+0.3cm,\y2);

\draw [arrow] (veriftest.west) -- node [pos=0.06,above] {No} node [pos=0.65, below]{Revise experiments or \\ calibration data or \\ DD approach} (notverif) -- (calverif);

\draw [arrow] (experiment) --  (Valexperiment);
\draw [arrow] (experiment)  -- node [pos=0.5,below] {UQ} (Calexperiment);

\coordinate (aboveDD) at ($(appsel.north)+(0cm,-1.8cm)$);
\draw [arrow,dashed,thick] (appsel.south) --node[pos=0.5, circle, inner sep=0pt, minimum size=3mm, draw,right=1mm,solid,thin]{+}  (aboveDD);

\coordinate (rightequi) at ($(equi.east)+(2.25cm,0cm)$);
\draw [arrow,dashed,thick] (equi.east) --node[pos=0.5, circle, inner sep=0pt, minimum size=3mm, draw,above=.5mm,solid,thin]{+} (rightequi);

\path [line2] (conceptModelT) -- (conceptModel.center);

\end{tikzpicture}
\end{center}
\caption{Flow chart of model development, verification and validation of a computational model involving DD constitutive models.} \label{fig:VandVChart}
\end{figure}
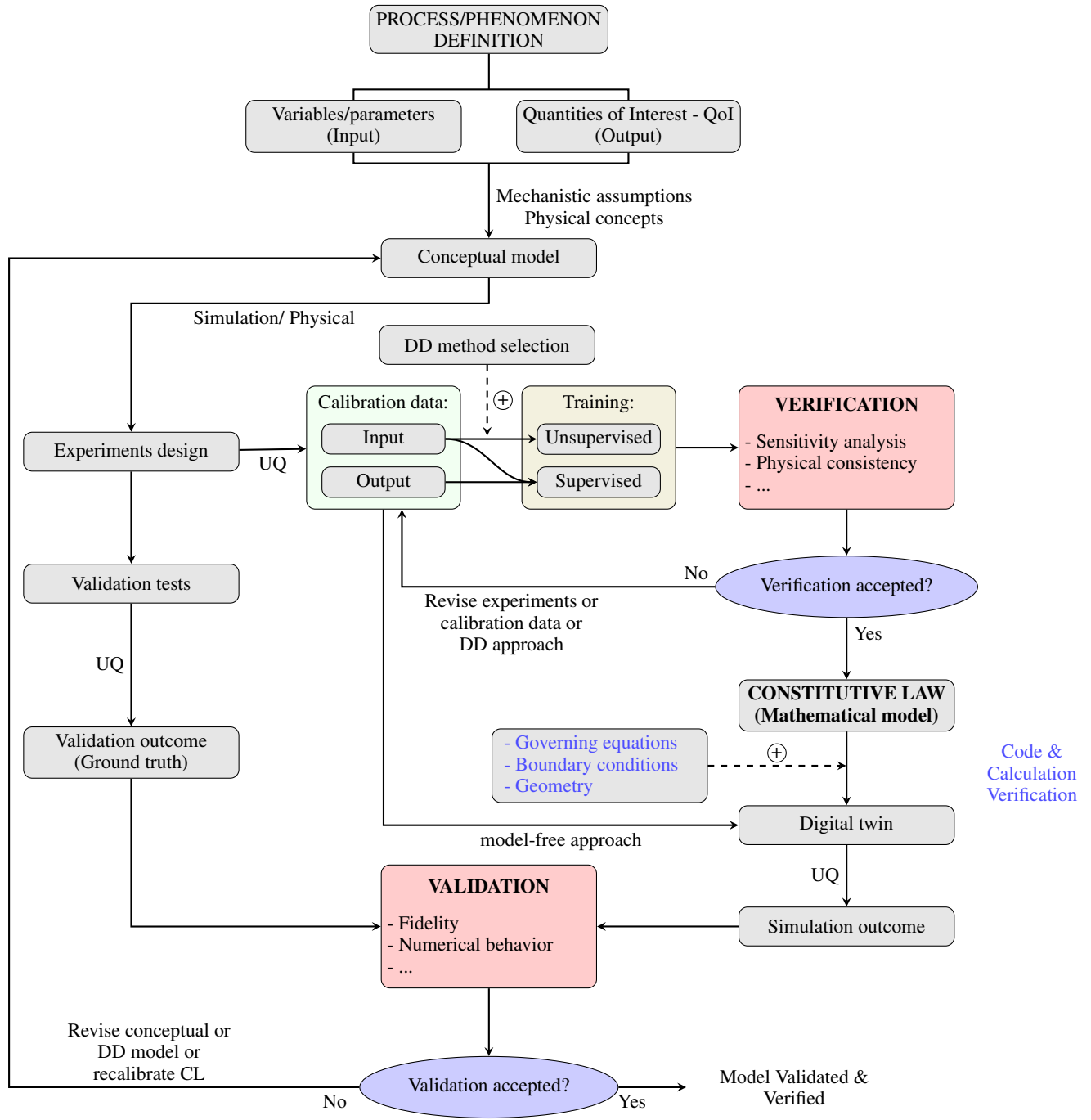

In the context of the present work, the V\&V process of DD constitutive modeling involves the same steps as those of traditional phenomenological constitutive models (c.f. \cite{roache1998verification,jones2008finite,henninger2010validation}), however, in light of big data availability, the possible black-box character and presence of many parameters of some ML approaches, different additional caveats have to be accounted for. 
The general workflow of the process is inspired by \cite{thacker2004concepts} 
and is schematized in Figure~\ref{fig:VandVChart}. 
The validation tests and their outcome including the UQ analysis constitute the ground truth obtained from the experimental side (which could come from a simulation or a physical experiment), while the right branch starting from the definition of the calibration data highlights the development and employment of the computational model. The general V\&V process starts with the definition of the process or phenomenon of interest and the identification of the relevant or available variables and parameters (which possibly constitute the \textit{inputs} of the CL) and the \textit{Quantities of Interest} (QoI) which we aim to predict (namely, the \textit{output} of the CL). Then, a \textit{conceptual model} has to be developed that includes the general objective, a target accuracy between modeled and true QoI, as well as necessary physical and mechanical assumptions of the model. In this phase, the independent governing variables whose effect should be included and investigated must be defined (e.g., which environmental parameters should be included as variables). Based on the conceptual model, an \textit{experiments design} should be carried out so as to allow for, on the one hand, the collection of the \textit{calibration data}, and on the other hand the definition of the validation tests. 

Once the input and output quantities (variables/parameters and QoIs) are collected and their UQ is performed, they can be adopted for the training of the selected DD approach, following a supervised (labeled inputs and outputs from the calibration data are used) or unsupervised approach (only the inputs are used). At this point, the trained DD approach becomes a candidate CL and it must be verified in order to assess if and how well it is able to represent the calibration data and the physics behind it given a set of desirable properties measured by a selection of performance metrics. In this context, the accuracy and sensitivity to the training data are of utmost importance as well as the physical consistency checks, the fidelity (given the task for which the CL is needed), and the reliability. Therefore, if the verification is not satisfied the operator must revise and improve all or some of the steps from the definition of the calibration data to the selection and training of the DD approach until the verification test is passed. At that moment, the approach (intended as a mathematical description of a set of data) can be considered representative of a material behavior and, hence, a CL.

Once the verified CL is complemented with governing equations, boundary (and, possibly, initial) conditions, and the geometry a \textit{digital twin} of the investigated process is created. Note that, in the case of a model-free approach the calibration data are directly injected into the digital twin while avoiding their manipulation. This also implies the absence of any CL in terms of a mathematical description. The digital twin can be used for the validation, which is assessed with respect to (at least) two aspects: (i) the capability of the digital twin to reproduce the behavior of real components (namely, the fidelity) and (ii) the numerical behavior of the model.  In the former case, the digital twin is adapted to the validation experiments by prescribing the proper geometry and boundary conditions and then the outcomes of the validation tests and of the digital twin simulations are quantitatively compared to determine if the trained CL is able to reproduce the phenomenon of interest. We remark that it is here particularly important to adopt independent validation tests, namely tests whose results cannot be represented by simple combinations of the calibration data. In other words, the validation data should also test the capacity of the CL to reproduce loading paths not included in the calibration dataset. Particular attention should be paid here in testing the possible overfitting of the data \cite{ying2019overview,ghojogh2019theory,buhlmann2011statistics}, an issue that can lead to inconsistent predictions of the material behavior. In this phase, it is also important to perform a UQ to estimate the confidence of the approach to replicate real behaviors and compare it to the one of the experiments. Concerning the numerical behavior it is important to evaluate how the DD approach integrates into the numerical solver (e.g., possibility of obtaining a consistent tangent matrix, smoothness of the operators) and the overall convergence behavior of the model. For both aspects, a set of metrics should be defined to assess the success or rejection of the validation procedure. In the negative case, the whole process should be revised, including the definition of the digital twin and the numerical framework thereof.

Even though less developed than in computational mechanics, V\&V in the context of DD model development has been pushed more into the focus lately \cite{ding2017machine,pei2017towards,xiang2018verification} due to the rising popularity of DD even in high-consequence applications.
In the following, we briefly revise the available contributions in the context of the V\&V of DD approaches. As a general comment, we remark that the available literature deals mostly with NNs, while other approaches are less investigated.

 \paragraph{DD verification.} 
 
 In principle verifying a DD model requires checking that the model is able to broadly represent its intended use case and does not behave unexpectedly (e.g., blow-up, undefined output for certain inputs) under a broad range of circumstances. Also, the operator should test the sensitivity of the model to both free parameters of the training process as well as calibration data quantity and quality. Due to the general opaqueness (black-box nature) of many DD models, corner cases are generally more unpredictable than for traditional phenomenological modeling. Therefore, for practitioners, it is crucial to answer the question as to which specific range of input values the model should be verified against. In principle, the latter should be sufficient to explore all the possible practical cases of interest, however, if this is not possible, the range of input tested should always be documented and made available.

Another common type of ML verification is the check for local robustness of a model, see e.g. \cite{huang2017safety} for a definition. Essentially a small perturbation in the input should only lead to a small change in the output.
Adversarial attacks \cite{carlini2019evaluating} are typically employed in the literature to provide evidence that the model is robust when no provable guarantee is available.
Other critical verification properties include the output reachability property, see e.g. \cite{ruan2018reachability}, i.e. if we know that a given stress component value can reach a certain magnitude we need to ensure that our ML output is able to represent this number.
For more information on and general definitions of verification techniques for ML models (specifically, fully connected NNs), we refer to \cite{huang2020survey}. In the context of model-free approaches, verification usually takes the form of a convergence study with respect to the number of points available in the material dataset and to the noise amplitude \cite{kirchdoerfer2016data,conti2018data,Carrara2020}. 
  
Verification of ML models has seen less attention in the literature compared to their validation.

 \paragraph{DD validation.}

Validation describes the process of assessing if an approach reproduces the phenomenon of interest in a broad range of applications and its predictive capabilities. Compared to classical ML tasks, this is especially important for constitutive modeling of solids due to the low/limited data availability often encountered in this area and the frequent need to extrapolate outside of the training domain. In the following we review only the aspects of validation that are specific to DD approaches for CLs; for the classical aspects (e.g., comparison with validation tests and convergence) we refer to the available literature   \cite{wallace1989software, thacker2004concepts,Zienkiewicz2005, Hughes2008}, which deals particularly with the issue of over-parametrization for the ML approaches involving many fitting parameters.

In contrast to phenomenological modeling, over-parametrization is a typical problem in ML constitutive modeling, hence, validating an ML model often requires checks for overfitting because it is a major sign of subpar generalization behavior.
Various strategies have been proposed in the literature to either detect overfitting or prevent it, see \cite{ying2019overview,ghojogh2019theory} for more information.
The modification of an ML algorithm so as to prevent overfitting is often referred to as regularization \cite{buhlmann2011statistics}. In DD constitutive modeling, the following explicit or implicit regularization methods have been proposed: 
    \begin{itemize}
\item Early stopping. The most common type of validation is built upon
validation datasets that can be used for implicit regularization by early stopping (stopping training when the error on the validation dataset increases, as this is a sign of overfitting to the training dataset) \cite{vabalas2019machine}. Splitting the dataset into training and validation components is not always straightforward especially if time- or history-dependent material behavior is studied \cite{bergmeir2012use}.
    
\item $L^{p}$ regularization. $L^{p}$ regularization in general adds a term to the training loss function that penalizes the complexity of the trainable parameters of the ML model.
Due to the reduction of model complexity, it is hoped to avoid overfitting and reduce the generalization error.
Common regularization techniques of this type include
 \textit{Ridge} regression or $L^{2}$-regularization \cite{hoerl1970ridge}, which is used to make the amplitude of the trainable parameters of a model smaller \cite{van2017l2}. In an adjusted form $L^{2}$-regularization can appear as "weight decay" in ML when using stochastic gradient descent schemes \cite{loshchilov2017decoupled}.
$L^{p}$-regression with $0< p\leq 1$ can also be used to
enforce sparsity on a model, which can reduce the number of trainable parameters and the sensitivity to noise while making the model more interpretable \cite{frank1993statistical,flaschel_unsupervised_2021,flaschel2022discovering}. 
The special case $p=1$ is known as \textit{Lasso} regression \cite{tibshirani1996regression}.
Other methods such as \textit{elastic net} \cite{zou2005regularization} use a combination of \textit{Ridge} and \textit{Lasso}. 
\item Dilution and Dropout. When training NNs, Dilution and Dropout are used to randomly drop units and relevant connections from the NNs during training. This prevents weights from co-adapting too much, thereby aiding generalization and avoiding overfitting \cite{srivastava2014dropout}. This technique can generally be employed to regularize any ML model and was e.g. employed by \cite{vlassis2020geometric} to obtain DD models for anisotropic hyperelasticity. 
    \item Physics-based regularization. 
    The generalization error of a DD model can be vastly improved by either implicitly or explicitly adding physics constraints that underlie the data to the model. In constitutive modeling, this can take the form of implicitly enforcing polyconvexity
    \cite{klein2022polyconvex,kalina2022automated} of hyperelastic laws, convexity of yield functions \cite{fuhg2022machine} or thermodynamic consistency of time- and history-dependent material models \cite{gonzalez2019thermodynamically,fuhg2023modular}.
    \item Adversarial ML. Originally developed to prevent model exploitation against attacks, adversarial training, i.e. training on intentionally misleading or perturbed data, can help the model to generalize better and become more robust \cite{stutz2019disentangling}. This includes schemes such as weight perturbation
    \cite{wu2020adversarial} or training on perturbed inputs which are commonly referred to as
adversarial examples \cite{raghunathan2019adversarial}.
    \cite{stocker2022novel} uses this technique to improve the generalization performance of an elastoplastic material model trained by an RNN. 
    \item Post-hoc explainability. Black-box models, e.g. those based on NNs, can be validated after the training process has ended by trying to produce useful approximations of the decision logic of the model that correspond to understandable/interpretable representations. This process typically involves the generation of a second (post-hoc) model that is used to explain the behavior of the first black-box model.
    This can e.g. be achieved by pruning the model \cite{blalock2020state}, e.g. using parameter reduction techniques, or training an interpretable model e.g. with symbolic or polynomial regression.
    The latter has e.g. been deployed in \cite{meyer2023} for modeling plasticity.
    \end{itemize}
Even though many different ML validation techniques have been introduced to counteract overfitting and help improve generalization, one of the open questions in the field is how ML tools can be certified \cite{torens2022machine}. There is a lack of general metrics that specify at which point an ML model can be trusted, i.e. how interpretable/trustworthy it needs to be. Furthermore, benchmarks have to be established to tell if a model can be deemed safe enough to be used even in high-risk applications.

\section{Conclusions}


 At this point, it is important to distill the main findings of this journey in constitutive modeling and to outline our view of the successes, open issues, and opportunities that lie ahead. Phenomenological constitutive approaches are by definition based on observations, making them in some sense the original DD models, but contrary to some of their more modern counterparts they are also interpretable, able to generalize, and not prone to overfitting. Borrowing some language from the ML community,  the archetype in mechanics for phenomenological constitutive modeling approaches has been to use simple experiments that result in homogeneous stress states (e.g. uniaxial, biaxial, simple shear) as ``training" data and to use structural responses that include inhomogeneous fields as ``validation". Even though these analytical approaches often lack expressivity, the strict enforcement of thermodynamic constraints, and balance laws, as well as the incorporation of mechanistic insight allow these models to extrapolate to unseen stress states and be useful and trustworthy for safety-critical predictive calculations in structural engineering as well as in other disciplines. As downsides of these successes, the development of classical phenomenological models requires specialized domain knowledge and a substantial conceptual effort; already the selection of the most suitable model among available options, along with its calibration, necessitates tedious and time-consuming trial-and-error procedures, which are a significant obstacle towards material innovation in many engineering fields.
 
 From our standpoint, DD constitutive modeling approaches, through the introduction of new computational tools, have the potential to: 1) accommodate the utilization of larger datasets from modern experimental techniques and assist in the interpretation of the data, 2) enhance the flexibility in the exploration and utilization of the modeling space to best interpret the data; 3) simplify the utilization of multifidelity and multimodal information (e.g. merge experimental and computational data), 4) automate the connection from data to predictive simulations streamlining the necessary infrastructure, 5) enable multiscale calculations, as well as 6) encode information for material variability as a first step towards uncertainty quantification and reliability analysis.

 As the second wave of DD constitutive modeling approaches is reaching a stage of maturation beyond initial exploratory attempts, it is becoming clear that a way to maintain the benefits of phenomenological modeling while developing and utilizing expressive and automated DD approaches is to take full advantage of a long and rich history of insightful mechanistic research formalized through what we have denoted throughout the paper as physics constraints. We have discussed several examples, all over different classes of DD constitutive models, where successes in terms of robustness, interpretability, generalization, and overall trustworthiness are gained by enforcing such constraints. 

 Even though some promising results have been recorded, there are still many open challenges. One of them is the seamless integration of DD approaches in the existing infrastructure of advanced experimental mechanics on the one hand, and of established computational engineering on the other hand. Experimental approaches based on modern imaging such as DIC or DVC rely on complex correlation algorithms that, starting from raw data such as grey level fields, infer full-field displacement information that can be directly utilized in learning tasks (e.g. DD discovery of interpretable CLs, or training of NN-based DD CLs). In principle, an integration between correlation and learning algorithms could reduce the accumulation of errors and streamline the process leading to the establishment of DD CLs from this type of experimental data. At the other end of the process, discovered or trained CLs need to be 
integrated with traditional computational mechanics tools such as finite element solvers, in order to be readily used by engineering practitioners in the solution of boundary value problems. Automatic differentiation, which is a key tool for this purpose, often leads to a decrease in computational performance over non-linear finite element solvers with hard-coded algorithms (e.g. in plasticity when using return mapping algorithms). 
 Note also that error measurements which commonly indicate the success of a DD approach are not sufficient to guarantee the convergence of non-linear finite element solvers embedding DD constitutive models. To this end, thorough testing of these models in a FEM setting is necessary.

 For path-dependent material behavior, 
an interesting task (and one that so far has not been sufficiently explored) is the discovery of underlying irreversible mechanisms and corresponding variables (e.g. hardening variables) directly from data. Also, sampling becomes significantly more complex and restricted from experimental protocols, as the learning task focuses on objects that depend on internal variables (e.g. yield function).

Additionally, we should strive for DD constitutive models to be robust in a low-data setting (contrary to the ongoing trends in computer science), as one could envision access to limited experimental samples and also limited computational resources, especially as multiscale problems become more complex and path-dependent behaviors are considered. In principle, already available DD approaches are able to discover CLs on a one-shot basis, i.e. using only one test. Clearly, the success of this approach needs a sufficiently complex test geometry and a sufficiently comprehensive test method (e.g. including loading and unloading stages, possibly at different rates). Thus, the optimal design of test methods and specimen geometries to guarantee identifiability (especially critical for highly expressive models) could deliver important contributions in this respect and is still largely unexplored.
In all these endeavors, UQ is expected to play a major role. Fundamental questions related e.g. to the amount and quality of data needed to deliver a DD model (or several competing models) of target uncertainty, and to the uncertainty of the outcomes of downstream tasks such as the solution of non-linear problems embedding DD CLs can only be answered in a probabilistic framework.
 
 Finally, tasks such as the curation of large datasets along with corresponding benchmark problems should be established as a way to evaluate constitutive DD approaches against the criteria that we have defined in this work. Specifically, here we suggest starting with two open datasets for DD constitutive approaches in the context of hyperelasticity. The first focuses on RVE-generated data (in the context of computational homogenization) for a composite microstructure sampled uniformly in a hypersphere of the components of the deformation gradient and also includes some specific loading paths up to larger deformations as a means to test generalization. To connect with experiments that do not produce labeled data pairs, the second dataset and benchmark should correspond to full-field displacement data and reaction forces for a specific geometry and boundary value problem for training (computationally generated through FEM), and a complementary dataset from another geometry and boundary value problem to validate the DD constitutive model. More work needs to be done on this front and the community can only benefit by establishing similar benchmarks for more complex, path-dependent problems.

\section*{Author contribution statement}
The authors confirm contribution to the paper as follows: ;JF, NB, WS, and LDL work conceptualization; NB, WS, and LDL funding acquisition; JF, NB, GAP, BB, WS, NNV, MF, PC and LDL draft manuscript preparation and editing; all authors reviewed the manuscript and approved submission.

\section*{Acknowledgements}
MF, PC and LDL acknowledge support from the Swiss National Science Foundation (SNF), project number 200021\_204316 “Unsupervised data-driven discovery of material laws”. JF and NB gratefully acknowledge support by the Air Force Office of Scientific Research under award number FA9550-22-1-0075.
GAP and NB were supported by the SciAI Center, and funded by the Office of Naval Research (ONR), under Grant Number N00014-23-1-2729.
WS, BB and NNV are supported by  the National Science Foundation under grant contracts CMMI-1846875 and the Department of Energy, National Nuclear Security Administration, Predictive Science Academic Alliance Program (PSAAP) under Award Number DE-NA0003962. 
\clearpage
\typeout{}

\bibliography{bib_unified}

\end{document}